\documentclass{tlp}

\usepackage{amsmath}
\usepackage{xspace}
\usepackage[compact]{titlesec}
\usepackage{pstricks}

\newcommand{\authornote}[3]{
}

\newcommand{\paul}[1]{\authornote{Paul}{blue}{#1}}
\newcommand{\peter}[1]{\authornote{Peter}{green}{#1}}
\newcommand{\zoltan}[1]{\authornote{Zoltan}{red}{#1}}

\newcommand{\code}[1]{{\tt#1}}

\newlength{\figboxwidth}
\newcommand{\iclp}[1]{{#1}}
\newcommand{\tr}[1]{}
\newcommand{\reviewcomments}[1]{}

\newcommand{\longex}[1]{
	\ifthenelse{
		\equal{\whichexample}{long}
	}{
		{#1}
	}{
		{}
	}
}
\newcommand{\shortex}[1]{
	\ifthenelse{
		\equal{\whichexample}{short}
	}{
		{#1}
	}{
		{}
	}
}

\newcommand{\picfigure}[2]{
	\begin{figure}[t]
	\newcommand{\spf}{\footnotesize}      %
	\input{pics/#1}                      %
	\centerline{\raise 1em\box\graph}     %
	\vspace{1mm}
	\caption{#2}
	\label{fig:#1}
	\end{figure}
}

\newcommand{\mapfoldl}{\code{map\_foldl}\xspace}

\newcommand{\todoitem}[3]{\parbox{4in}{#1} & 
  \parbox{1in}{#2} & \parbox{1in}{#3} }

\newcommand{\squishlist}{
  \begin{list}{$\bullet$} {
    \setlength{\itemsep}{0pt}
    \setlength{\parsep}{3pt}
    \setlength{\topsep}{3pt}
    \setlength{\partopsep}{0pt}
    \setlength{\leftmargin}{1.5em}
    \setlength{\labelwidth}{1em}
    \setlength{\labelsep}{0.5em}
  }
}

\newcommand{\squishend}{
  \end{list}
}

\title[Estimating the overlap between dependent computations]
{Estimating the overlap between dependent computations
for automatic parallelization}

\author[Paul Bone, Zoltan Somogyi and Peter Schachte]
    {PAUL BONE\thanks{work supported by an Australian Postgraduate Award
    and a NICTA top-up scholarship.},
    ZOLTAN SOMOGYI\\
           Department of Computer Science and Software Engineering \\
           The University of Melbourne and \\
           National ICT Australia (NICTA) \\
           \email{\{pbone,zs\}@csse.unimelb.edu.au}
    \and PETER SCHACHTE\\
           Department of Computer Science and Software Engineering \\
           The University of Melbourne \\
           \email{schachte@unimelb.edu.au}
    }

\begin{document}

\maketitle

\begin{abstract}
Researchers working on the automatic parallelization of programs
have long known that too much parallelism
can be even worse for performance than too little,
because spawning a task to be run on another CPU incurs overheads.
Autoparallelizing compilers have therefore
long tried to use granularity analysis
to ensure that they only spawn off computations
whose cost will probably exceed the spawn-off cost by a comfortable margin.
However, this is not enough to yield good results,
because data dependencies may \emph{also} limit
the usefulness of running computations in parallel.
If one computation blocks almost immediately
and can resume only after another has completed its work,
then the cost of parallelization again exceeds the benefit.

We present a set of algorithms for recognizing places in a program
where it is worthwhile to execute two or more computations in parallel
that pay attention to the second of these issues as well as the first.
Our system uses profiling information to compute
the times at which a procedure call consumes the values of its input arguments
and the times at which it produces the values of its output arguments.
Given two calls that may be executed in parallel,
our system uses the times of production and consumption
of the variables they share
to determine how much their executions would overlap
if they were run in parallel,
and therefore whether executing them in parallel is a good idea or not.

We have implemented this technique for Mercury
in the form of a tool that uses profiling data
to generate recommendations about what to parallelize,
for the Mercury compiler to apply on the next compilation of the program.
We present preliminary results that show that
this technique can yield useful parallelization speedups,
while requiring nothing more from the programmer
than representative input data for the profiling run.
\end{abstract}

\begin{keywords}
automatic parallelism,
program analysis,
program optimization,
Mercury
\end{keywords}

\section{Introduction}
\label{sec:intro}

\begin{figure}[tb]
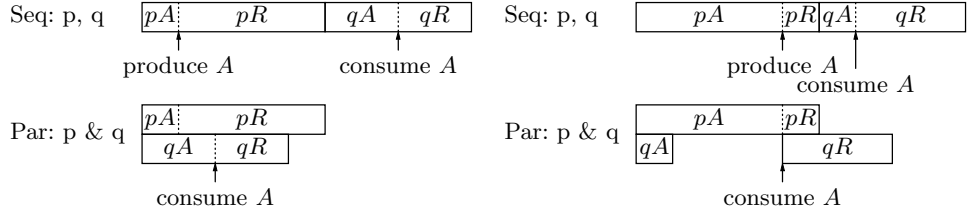

\figrule
\vspace{-2\baselineskip}
\begin{center}
\include{small_overlap}
\centerline{\raise 1em\box\graph}
\end{center}
\caption{Ample vs smaller parallel overlap between \code{p} and \code{q}}
\label{fig:dep_conj_overlap1}
\figrule
\vspace{-2\baselineskip}
\end{figure}

When parallelizing Mercury~\cite{jlp} programs,
the best parallelization opportunities occur
where two goals take a significant and roughly similar time to execute.
Their execution time should be as large as possible
so that the relative costs of parallel execution are small,
and they should be independent to minimize synchronization costs.
Unfortunately, goals expensive enough to be worth executing in parallel
are rarely independent.
For example, in the Mercury compiler itself,
there are 53 conjunctions containing two or more expensive goals,
but in only one of those conjunctions are the expensive goals independent.
This is why Mercury supports the parallel execution of dependent conjunctions.
The Mercury compiler wraps shared variables within a
\emph{future}~\cite{wang_dep_par_conj}, to
ensure that the \emph{consumer} of the variable is blocked
until the \emph{producer} makes the variable available.

Dependent parallel conjunctions may differ
in the amount of parallelism they have available.
Consider a parallel conjunction with two similarly-sized conjuncts,
\code{p} and \code{q}, that share a single variable \code{A}.
If \code{p} produces \code{A} late but \code{q} consumes it early,
as shown on the right side of figure \ref{fig:dep_conj_overlap1},
there will be little parallelism,
since \code{q} will be blocked soon after it starts,
and will be unblocked only when \code{p} is about to finish.
Alternatively, if \code{p} produces \code{A} early
and \code{q} consumes it late,
as shown on the left side of in figure~\ref{fig:dep_conj_overlap1},
we would get much more parallelism.
The top part of each scenario
shows the execution of the sequential form of the conjunction.

Unfortunately, in real Mercury programs,
almost all conjunctions are dependent conjunctions,
and in most of them,
shared variables are produced very late and consumed very early.
Parallelizing them would therefore yield slowdowns instead of speedups,
because the overheads of parallel execution would far outweigh the benefits.
We want to parallelize only conjunctions
in which any shared variables are produced early, consumed late,
or (preferably) both.
The first purpose of this paper is to show how one can find these conjunctions.

\begin{figure}[b]
\figrule
\begin{verbatim}
map_foldl(_, _, [], Acc, Acc).
map_foldl(M, F, [X | Xs], Acc0, Acc) :-
    M(X, Y),
    F(Y, Acc0, Acc1),
    map_foldl(M, F, Xs, Acc1, Acc).
\end{verbatim}
\caption{\mapfoldl}
\label{fig:map_foldl}
\figrule
\vspace{-2\baselineskip}
\end{figure}

The second purpose is to find the best way to parallelize these conjunctions.
Consider the \mapfoldl predicate in figure~\ref{fig:map_foldl}.
The body of the recursive clause has three conjuncts.
We could make each conjunct execute in parallel,
or we could execute two conjuncts in sequence
(either the first and second, or the second and the third),
and execute that sequential conjunction in parallel with the remaining conjunct.
In this case, there is little point in executing
the higher order calls to the map and fold predicates
in parallel with one another,
since in virtually all cases,
the map predicate will generate \code{Y} very late and
the fold predicate will consume \code{Y} very early.
However, executing the sequential conjunction of the map and fold predicates
in parallel with the recursive call \emph{will} be worthwhile
if the map predicate is time-consuming,
because this implies that
a typical recursive call will consume its fourth argument late;
the recursive call processing the second element of the list
will have significant execution overlap
with its parent processing the first element of the list
even if (as is typical) the fold predicate generates \code{Acc1} very late.
(This is the kind of computation that
Reform Prolog \cite{bevemyr:reform} was designed to parallelize.)

The structure of this paper is as follows.
Section~\ref{sec:background} gives
the background needed for the rest of the paper.
Section~\ref{sec:approach} outlines our general approach,
which the later sections fill in.
Section~\ref{sec:overlap} describes our algorithm for calculating
the execution overlap between two or more dependent conjuncts.
A conjunction with more than two conjuncts can be parallelized
in several different ways;
section~\ref{sec:howto} shows how we choose the best way.
\tr{Section~\ref{sec:pragmatic} discusses some pragmatic issues.}
Section~\ref{sec:perf} evaluates
how our system works in practice on some example programs, and
section~\ref{sec:conc} concludes
with comparisons to related work.

\section{Background}
\label{sec:background}

\subsection{Mercury}
\label{sec:backmer}

\tr{
Mercury is a pure logic/functional programming language
intended for the creation of large, fast, reliable programs.
While the syntax of Mercury is based on the syntax of Prolog,
semantically the two languages are very different
due to Mercury's purity and its type, mode, determinism and module systems.
}

The abstract syntax of the part of Mercury relevant to this paper is:

\vspace{-1\baselineskip}
$$
\begin{array}{lll}
\mbox{pred}~P
    & :~ p(x_1, \ldots, x_n)~\leftarrow~G
        & \hbox{predicates} \\
\mbox{goal}~G
    & :~ x = y ~|~ x = f(y_1,~\ldots,~y_n)
        & \hbox{unifications}\\
    & |~ p(x_1,~\ldots,~x_n) ~|~ x_0(x_1,~\ldots,~x_n)
        & \hbox{first and higher order calls} \\
    & |~ (G_1,~\ldots,~G_n) ~|~ (G_1~\&~\ldots~\&~G_n)
        & \hbox{seq and par conjunctions}\\
    & |~ (G_1 ; \ldots ; G_n) ~|~ \hbox{switch}~x~(\ldots;~f_i: G_i ; \ldots)
        & \hbox{disjunctions and switches}\\
    & |~ (if~G_c~then~G_t~else~G_e) ~|~ not~G
        & \hbox{if-then-elses and negations}\\
    & |~ some~[x_1,\ldots,x_n]~G
        & \hbox{quantifications}\\
\end{array}
$$
\label{fig:abstractsyntax}
\vspace{-1mm}

\noindent
The atomic constructs of Mercury are unifications
(which the compiler breaks down until they contain
at most one function symbol each),
plain first-order calls,
and higher-order calls.
The composite constructs include
sequential and parallel conjunctions,
disjunctions, if-then-elses, negations and existential quantifications.
These should all be self-explanatory.
A switch is a disjunction in which
each disjunct unifies the same bound variable
with a different function symbol.
\tr{
Switches in Mercury are thus analogous to switches in languages like C.
}

\tr{
Mercury has a strong Hindley-Milner type system very similar to Haskell's.
Mercury programs are statically typed;
the compiler knows the type of every argument of every predicate
(from declarations or inference) and every local variable (from inference).
}

Mercury \tr{also} has a strong mode system.
The mode system classifies each argument of each predicate
as either input or output;
there are exceptions, but they are not relevant to this paper.
If input, the caller must pass a ground term as the argument.
If output, the caller must pass a distinct free variable,
which the predicate\tr{ or function} will instantiate to a ground term.
It is possible for a predicate\tr{ or function} to have more than one mode;
we call each mode of a predicate\tr{ or function} a \emph{procedure}.
The \tr{Mercury} compiler generates separate code
for each procedure of a predicate\tr{ or function}.
The mode checking pass of the compiler is responsible for
reordering conjuncts (in both sequential and parallel conjunctions)
as necessary to ensure that for each variable shared between conjuncts,
the goal that generates the value of the variable (the \emph{producer})
comes before all goals that use this value (the \emph{consumers}).
This means that for each variable in each procedure,
the compiler knows exactly where that variable gets grounded.

Each procedure and goal has a determinism,
which may put upper and lower bounds on the number of its possible solutions
(in the absence of infinite loops and exceptions).
A determinism may impose an upper bound of one solution,
and it may impose a lower bound of \tr{either zero solutions or} one solution.
\emph{det} procedures succeed exactly once;
\tr{(upper bound is one, lower bound is one);}
\emph{semidet} procedures succeed at most once;
\tr{(upper bound is one, no lower bound);}
\emph{multi} procedures succeed at least once;
\tr{(lower bound is one, no upper bound);}
\emph{nondet} procedures may succeed any number of times\iclp{.}
\tr{(no bound of either kind).}
\tr{
Goals with determinism \emph{failure} can never succeed
(upper bound is zero, no lower bound).
Goals with determinism \emph{erroneous}
have an upper bound of zero and a lower bound of one,
which means they can neither succeed nor fail,
so the only things they can do is throw an exception or loop forever.
}

\tr{
The compiler keeps a lot of information associated with each goal,
whether atomic or not.
This includes:
\vspace{3mm}
\begin{itemize}
\item
the set of variables bound (or \emph{produced}) by the goal;
\item
the \emph{nonlocal set} of the goal,
which means the set of variables
that occur both inside the goal and outside it; and
\item
the determinism of the goal.
\end{itemize}
\vspace{3mm}
}

\subsection{Parallelism in Mercury}
\label{sec:backpar}

The Mercury runtime system has a construct called a Mercury \emph{engine}
that represents a virtual CPU.
Each engine is independently schedulable by the OS, usually as a POSIX thread.
The number of engines that a parallel Mercury program will allocate on startup
is configurable by the user,
but it defaults to the actual number of CPUs.
Another construct in the Mercury runtime system is a \emph{context},
which represents a computation in progress.
An engine may be idle, or it may be executing a context;
a context can be running on an engine, or it may be suspended.
When a context finishes execution,
its storage is put back into a pool of free contexts.
Following \citeN{simonmar_2009_multicore_rts},
we use \emph{sparks} to represent goals that have been spawned off
but whose execution has not yet been started.
\tr{
when an engine executes a spark, it XXX users its current context,
or if it does not have a context a new one is allocated
(from the pool of free contexts if the pool is not empty,
and a newly created context otherwise).
Unlike \cite{simonmar_2009_multicore_rts}, Mercury's sparks cannot be
garbage collected and must be executed.
}

The only parallel construct in Mercury is parallel conjunction,
which is denoted $(G_1~\&~\ldots~\&~G_n)$.
All the conjuncts must be deterministic,
that is, they must all have exactly one solution.
This restriction greatly simplifies the implementation,
since it guarantees that there can never be any need
to execute $(G_2~\&~\ldots~\&~G_n)$ multiple times,
just because $G_1$ has succeeded multiple times.
(Any local backtracking inside $G_1$ will not be visible to the other conjuncts;
bindings made by det code are never retracted.)
However, this is not a significant limitation.
Since the design of Mercury strongly encourages deterministic code,
in our experience, about 75 to 85\% of all Mercury procedures are det,
and most programs spend an even greater fraction of their time in det code.
Existing algorithms for executing nondeterministic code in parallel
have very significant overheads, generating slowdowns by integer factors.
Thus we have given priority to parallelizing deterministic code,
which we can do with \emph{much} lower overhead.

The Mercury compiler implements $(G_1~\&~G_2~\&~\ldots~\&~G_n)$
by creating a data structure representing a barrier,
and then spawning off $(G_2~\&~\ldots~\&~G_n)$ as a spark.
\tr{
The spark is added to the head of the context's double-ended queue.
If the context finishes the execution of the first conjunct
and finds that spark is still at the head of its queue,
it will pick it up and run it itself.
This is a useful optimization,
since it avoids using a separate context in the relatively common case
that all the other CPUs are busy with their own work.
(Contexts contain stacks, so they occupy nontrivial amounts of memory.)
}
Since $(G_2~\&~\ldots~\&~G_n)$ is itself a conjunction,
it is handled the same way:
the context executing it
first spawns off $(G_3~\&~\ldots~\&~G_n)$, and then executes $G_2$ itself.
Eventually, the spawned-off remainder of the conjunction
consists only of the final conjunct, $G_n$,
and the context just executes it.
\iclp{
The code of each conjunct synchronizes on the barrier once it has 
completed its job.
When all conjuncts have done so,
the original context will continue execution after the parallel conjunction.
}

\tr{
When an engine becomes idle, it will first try
to resume a suspended but runnable context if there is one.
If not, it will attempt to \emph{steal} sparks
from the tails of another context's queues.
If successful, it will allocate a context,
and start running the spark in that context.
The work-stealing dequeue structure we are using
is from \cite{Chase_2005_wsdeque}.
}

\tr{
After the end of the code of each conjunct,
the compiler inserts synchronization code.
This code starts by decrementing the number of outstanding conjuncts
in the conjunction's syncterm,
a number that was initialized to the number of conjuncts.
The engine executing the context that created the syncterm
will check whether there are any outstanding conjuncts in the conjunction.
If not, it will jump to the code after the conjunction.
If there are some,
it will suspend its context and look for other work to do.
An engine executing any other context will also check
whether there are any outstanding conjuncts when they finish.
If there are some, it will just look for other work.
If there are none, then it will first wake up the original context,
which must still be suspended.
The context it switches to executing after that
may or may not be the now runnable syncterm-creating context.
We have a special provision to avoid the potential race-condition
between one engine suspending the original context
at the same time as another engine is trying to wake it up.
}

\tr{
If an engine with a context other than the original one finds that
there are still outstanding jobs, they check their local spark queue
for any such work, otherwise they look for global work.

An engine looks for global work first by checking the local spark
queue of its context.
In some cases this check can be optimized out, for example after there
are no outstanding conjuncts in a parallel conjunction.
It will then check for runnable but suspended contexts,
If it finds a runnable context and is still holding a context from a
previous execution, it saves the old context onto the free context list.
If there are no runnable contexts,
it will then attempt to steal work from other contexts.
If unsuccessful, it will become idle and sleep for a period of time
before it looks for work again
or is woken up because a context has become runnable.
}

Mercury's mode system allows a parallel conjunct to consume variables
that are produced by conjuncts to its left, but not to its right.
This guarantees the absence of circular dependencies
and hence the absence of deadlocks between the conjuncts,
but it does allow a conjunct to depend on data that is yet to be computed
by a conjunct running in parallel.
We handle these dependencies through a source-to-source transform
\cite{wang_dep_par_conj}.
The compiler knows which variables
are produced by one parallel conjunct and consumed by another.
\iclp{
For each of these shared variables,
it creates a data structure called a \emph{future} \cite{multilisp}.
When the producer has finished computing the value of the variable,
it puts the value in the future and signals its availability.
When a consumer needs the value of the variable,
it waits for this signal,
and then retrieves the value from the future.
}
\tr{
For each of these shared variables,
the compiler creates a data structure called a \emph{future} \cite{multilisp},
which contains room for the value of the variable,
a flag indicating whether the variable has been produced yet,
a queue of consumer contexts waiting for the value, and a mutex.
The initial value of the future has the flag set to `not yet produced'.
The signal operation on the future sets the value of the variable,
sets the flag to `produced',
and wakes up all the waiting consumers,
all under the protection of the mutex.
The wait operation on the future is also protected by the mutex:
it checks the value of the flag,
and if it says `not yet produced',
the engine will put its context on the queue and suspend it before
looking for other work.
When it wakes up,
or if the flag said that the value was already `produced',
the wait operation simply gets the value of the variable.
}

To minimize waiting,
the compiler pushes signal operations
as far to the left into the producer conjunct as possible,
and it pushes wait operations
as far to the right into each of the consumer conjuncts as possible.
This means not only pushing them
into the body of the predicate called by the conjunct,
but also into the bodies of the predicates they call,
with the intention being that
each signal is put immediately after
the primitive goal that produces the value of the variable,
and each wait is put immediately before
the leftmost primitive goal that consumes the value of the variable.
Since the compiler has complete information
about which goals produce and consume which variables,
the only things that can stop the pushing process are
higher order calls and module boundaries:
the compiler cannot push a wait or signal operation
into code it cannot identify or cannot access.

\tr{
\subsection{The Mercury deep profiler}
\label{sec:backdeep}

The Mercury deep profiler~\cite{conway:2001:mercury-deep}
gathers much more information about the context of each measurement
than traditional profilers like \code{gprof}\cite{gprof}.
When it records the occurrence
of a call, a memory allocation or a profiling clock interrupt,
it records with it the chain of ancestor cliques
\paul{SCCs}
(groups of mutually recursive procedures)
all the way from the current call to the entry point of the program,
a procedure named \code{main}.
This detail allows the deep profiler
to find and present to the user
not just information such as the total number of calls to a procedure
and the average cost of a call,
or even information such as the total number of calls to a procedure
from a particular call site and the average cost of a call from that call site,
but also information such as the total number of calls to a procedure
\emph{from a particular call site
when invoked from a particular chain of ancestor cliques}
and the average cost of a call \emph{in that context}.
For example, it could tell that
procedure \code{h} called procedure \code{i} ten times
when \code{h}'s chain of ancestors was \code{main -> f -> h},
while \code{h} called \code{i} only seven times
when \code{h}'s chain of ancestors was \code{main -> g -> h},
the calls from \code{h} to \code{i} took on average twice as long
from the \code{main -> g -> h} context as from \code{main -> f -> h},
so that despite the fewer calls,
\code{main -> g -> h -> i} took more time than \code{main -> f -> h -> i}.
\paul{Reinforce why this is useful for auto-parallelism.
The reader may want to know why they're being given this information.}

Profilers have traditionally measured time
by sampling the program counter at clock interrupts.
Unfortunately, even on modern machines
the usual infrastructure for clock interrupts (\emph{e.g}., SIGPROF on Unix)
supports only one frequency for such interrupts,
which is usually 60 or 100 Hz.
This frequency is far too low for the kind of detailed measurements
the Mercury deep profiler wants to make,
since for typical program runs of few seconds,
it results in almost all calls having a recorded time of zero,
with the calls recording a nonzero time
(signifying the occurrence of an interrupt during their execution)
being selected almost at random.

We have therefore implemented a finer-grained measure of time
that turned out to be very useful
even though it is inherently approximate.
This measure is call sequence counts or CSCs:
the profiled program basically behaves
as if the occurrence of a call signified
the occurrence of a new kind of profiling interrupt.
In imperative programs, this would be a horrible measure,
since calls to different functions can have hugely different runtimes.
However, in declarative language like Mercury there are no explicit loops;
what a programmer would do with a loop in an imperative language
must be done by a recursive call.
This means that the only thing that the program can execute between two calls
is a sequence of primitive operations such as unifications and arithmetic.
For any given program,
there is a strict upper bound on the maximum length of such sequences,
and the distribution of the length of such sequences
is very strongly biased towards very short sequences
of half-a-dozen to a dozen operations.
In practice, we have found that
the fluctuations between the lengths of different such sequences
can be ignored for any measurement
that covers any significant number of call sequence counts (CSCs),
say more than a hundred.
The only drawback of this scheme that we have found
is that on 32 bit platforms,
its usability is limited to short program runs (a few seconds)
by the wraparound of the global CSC counter;
on 64 bit platforms, the problem would occur
only on a profiling run that lasts for years.
}

\section{Our general approach}
\label{sec:approach}

We want to find the conjunctions in the program
whose parallelization would be the most profitable.
This means finding the conjunctions with conjuncts
whose execution cost exceeds the spawning-off cost by the highest margin,
and whose interdependencies, if any,
allow their executions to overlap the most.
Essentially, the greater the margin by which
the likely runtime of the parallel version of a conjunction beats
the likely runtime of the sequential version,
the more beneficial parallelizing that conjunction will be.

To compute this likely benefit,
we need information
both about the likely cost of calls
and the execution overlap allowed by their dependencies.
Our system therefore asks programmers
to follow this sequence of actions
after they have tested and debugged the program.

\begin{enumerate}
\item
Compile the program
with options asking for profiling.
\item
Run the program on a representative set of input data.
This will generate a profiling data file.
\item
Invoke our feedback tool on the profiling data file.
This will generate a parallelization advice file.
\item
Compile the program for parallel execution,
specifying the parallelization advice file.
The advice file tells the compiler
\emph{which} sequential conjunctions to convert to parallel conjunctions,
and exactly \emph{how}.
For example, \code{c1, c2, c3} can be converted
into \code{c1 \& (c2, c3)},
into \code{(c1, c2) \& c3}, or
into \code{c1 \& c2 \& c3},
and as the \code{map\_foldl} example shows,
the speedups you get from them can be strikingly different.
\end{enumerate}

\noindent
It is up to the programmer using our system
to select training input for the profiling run in step 2.
Obviously, programmers should pick input that is as representative as possible,
but the recommended parallelization can be useful
even for input data that is quite different from the training input.
The main focus of this paper is on step 3;
we give the main algorithms used by the feedback tool.
\tr{
However, we will also touch on steps 1 and 4.
}

Our feedback tool is an extension of the Mercury deep profiler.
One of our modifications gives the deep profiler access
to the relevant parts of the compiler's representation of the program.
This includes a representation of each procedure body,
and for each atomic subgoal (call or unification) within each body,
the set of variables bound by that subgoal.
Another modification records
how many times execution reaches each point in the program.
\tr{
Since even the unmodified deep profiler could figure this out
for \emph{most} program points from the call counts associated with call sites,
we need to gather execution counts at only a few additional sites
to allow us to figure it out for \emph{all} program points.
}
As we will see in section~\ref{sec:overlap},
we need this information
to calculate the likely speedup from parallelizing a conjunction.

Our feedback tool looks for parallelization opportunities
by doing a depth-first search
of the call tree recorded in the profiling data file.
It explores the subtree below a node in the call tree
only if the overall cost of the call
is greater than a configurable threshold,
and if the amount of parallelism it has found at and above that node
is below another configurable threshold.
The first test lets us avoid looking at code
that would take more work to spawn off than to execute,
while the second test lets us avoid creating
more parallel work than the target machine can handle.

For each procedure in the call tree,
we search its body for conjunctions that contain
two or more calls with execution times above a configurable threshold.
To parallelize the conjunction,
its conjuncts have to be partitioned,
each partition being one conjunct in the parallel conjunction.
In most cases, this can be done in several different ways.
We can use the algorithms of section \ref{sec:overlap}
to compute the expected parallel execution time of each partition;
these algorithms take into account the runtime overheads of parallel execution.
We use the algorithms of section \ref{sec:howto} to generate
the set of partitions whose performance we want to evaluate.
If the best-performing parallelization we find
shows a nontrivial speedup over sequential execution,
we remember that we want to perform that parallelization on this conjunction.
If the depth first search later finds
some of the conjuncts to have parallelizable code inside them,
we revisit this conjunction,
this time using updated data about the cost of those conjuncts.
Otherwise,
we add a recommendation to perform the selected parallelization
to the feedback advice we generate for the compiler.

An important benefit of profile-directed parallelization is that
since programmers do not annotate the source program,
it can be re-parallelized easily after a change to the program
obsoletes some old parallelization opportunities and creates others.
Nevertheless, if programmers want to parallelize some conjunctions manually,
they can do so: our system will not override the programmer.

\section{Calculating the overlap between dependent conjuncts}
\label{sec:overlap}

As we can see from the difference between the two sides of
figure~\ref{fig:dep_conj_overlap1},
figuring out the overlap
in the parallel executions of two dependent conjuncts
requires knowing, for each of the variables they share,
when that variable is generated by the first conjunct and
when it is first consumed by the second conjunct.
Our algorithms for computing these times are considerably simplified
by the Mercury mode system
and by the fact that we only parallelize deterministic goals.

The profiling data gives us both
the total execution time of each conjunct
and its number of invocations;
the ratio of the two is the expected execution time for each invocation.
The algorithm for computing the expected production time
of a given shared variable looks at the form of the conjunct:
\begin{itemize}
\item
If the goal is a unification,
the expected production time is zero,
because our unit of time is the time between two successive calls.
\tr{
\item
If the goal is a foreign code goal,
the expected production time is the total cost of the goal.
Unless the foreign code calls back into Mercury,
we will measure this as exactly one CSC.
}
\item
If the goal is a first order call,
we recurse on the body of the callee.
\item
If the goal is a higher order call,
the expected production time is the cost of the call,
because the compiler cannot (yet) insert
the signalling of the future into the callee's body.
\item
If the goal is a conjunction $G_1,~\ldots,~G_n$,
and the variable is generated by $G_k$,
then we add up the total time taken by $G_1,~\ldots,~G_{k-1}$,
and add the sum to the result of invoking the algorithm recursively on $G_k$.
\item
If the goal is a switch,
we invoke the algorithm recursively on each switch arm,
and compute a weighted average of the results,
with the weights being the arms' entry counts.
\item
If the goal is an if-then-else,
we need the weighted average of the two possible cases:
the variable being generated by the then arm versus the else arm.
(It cannot be generated by the condition:
variables generated by the condition are visible only from the then-arm.)
To find the first number,
we invoke the algorithm on the then-arm,
and add the result to the time taken by the condition.
To find the second,
we invoke the algorithm on the else-arm,
and add the result to the expected time taken by the condition when it fails.
To compute this, we use a version of this algorithm
that weights the time taken by each conjunct in any inner conjunction
by the probability of its execution,
which we know by comparing its execution count
with the count of the number of times the condition was entered.
\item
The goal cannot be a negation, because negated goals cannot bind variables.
\item
The goal cannot be a disjunction, because
disjunctions cannot produce variables visible from det code.
(To transition from nondet or multi code to det code,
the programmer must quantify away the outputs of the nondet code.)
\item
If the goal is a quantification,
then the inner goal must be det,
in which case we invoke the algorithm recursively on it.
If the inner goal were not det,
then the outer quantification goal could be det
only if the inner goal did not bind any variables visible from the outside.
\end{itemize}

\noindent
Using the weighted average for switches and if-then-elses is meaningful because
the Mercury mode system dictates
that if one arm of a switch or if-then-else generates a variable,
then they \emph{all} must do so.
\tr{
The sole exception is arms that are guaranteed to abort the program,
whose determinism is erroneous.
We use a weight of zero for erroneous arms.
}

The algorithm we use for computing the time
at which a shared variable is first consumed by the second conjunct
is similar to this one,
the main differences being that
negated goals, conditions and disjunctions are allowed to consume variables,
and some arms of a switch or if-then-else
may consume a variable even if other arms do not.
Suppose the first appearance of the variable (call it $X$)
in a conjunction $G_1, \ldots, G_n$ is in $G_k$, and $G_k$ is a switch.
If $X$ is consumed by some switch arms and not others,
then on some execution paths,
the first consumption of the variable may be in $G_k$ (a),
on some others it may be in $G_{k_1}, \ldots, G_n$ (b),
and on some others it may not be consumed at all (c).
For case (a),
we compute the average time of first consumption by the consuming arms,
and then compute the weighted average of these times,
with the weights being the probability of entry into each arm, as before.
For case (b), we compute the probability of entry
into arms which do \emph{not} consume the variable,
and multiply the sum of those probabilities
by the weighted average of those arms' execution time
\emph{plus} the expected consumption time of the variable
in $G_{k+1},~\ldots,~G_n$.
For case (c)
we pretend $X$ is consumed at the very end of the goal,
and then handle it in the same way as (b).
This is because for our overlap calculations,
a goal that does not consume a variable is equivalent to
a goal that consumes it at the end of its execution.

Suppose a candidate parallel conjunction has two conjuncts $p$ and $q$,
and their execution times in the original, sequential conjunction $p, q$,
are ${SeqTime}_p$ and ${SeqTime}_q$.
Suppose ${SV}_i$ are the variables shared between them,
and for each ${SV}_i$,
the time at which $p$ produces it is ${ProdTime}_{pi}$, and
the time at which $q$ consumes it is ${ConsTime}_{qi}$.

If we denote the execution times of the conjuncts
in the parallel conjunction $p~\&~q$
as ${ParTime}_p$ and ${ParTime}_q$,
then the expected speedup
from parallelizing the original sequential conjunction
is ${Speedup} = {SeqTime} / {ParTime}$,
where ${SeqTime} = {SeqTime}_p + {SeqTime}_q$,
and ${ParTime} = {SpawnOverhead} + {max}({ParTime}_p, {ParTime}_q)$.
The profile gives us ${SeqTime}_p$ and ${SeqTime}_q$,
and if we ignore overheads for now (we will come back to them later),
then ${ParTime}_p$ will always be equal to ${SeqTime}_p$.
The main task of computing the speedup
therefore consists of computing ${ParTime}_q$;
as we saw in figure~\ref{fig:dep_conj_overlap1},
this will differ from ${SeqTime}_q$
whenever $q$ needs to wait for $p$ to produce a shared variable.

\begin{figure}[tb]
\figrule
\begin{center}
\begin{verbatim}
find_par_time(Conjs) returns TotalParTime:
  N := length(Conjs)
  ProdTimeMap := empty
  TotalParTime := 0
  for i in 1 to N:
    CurSeqTime := 0
    CurParTime := 0
    sort ProdConsList_i on Time_ij
    forall (Var_ij, Time_ij) in ProdConsList_i:
      Duration_ij := Time_ij - CurSeqTime
      CurSeqTime := CurSeqTime + Duration_ij
      if Conj_i produces Var_ij:
        CurParTime := CurParTime + Duration_ij
        ProdTimeMap[Var_ij] := CurParTime
      else Conj_i must consume Var_ij:
        ParWantTime := CurParTime + Duration_ij
        CurParTime := max(ParWantTime, ProdTimeMap[Var])
    DurationRest_i := SeqTime_i - CurSeqTime
    CurParTime := CurParTime + DurationRest_i
    TotalParTime := max(TotalParTime, CurParTime)
\end{verbatim}
\end{center}
\caption{Dependent parallel conjunction algorithm}
\label{fig:dep_par_conj_overlap_middle}
\figrule
\vspace{-2\baselineskip}
\end{figure}

Figure~\ref{fig:dep_par_conj_overlap_middle} shows
a simplified version of the algorithm we use to compute
the expected execution time of a conjunction
when its conjuncts are executed in parallel,
assuming an unlimited number of CPUs.
The inputs of the algorithm are \verb|Conjs|, the conjuncts themselves,
and \verb|ProdConsList|,
which gives, for each conjunct,
the list of its input and output variables,
together with the times at which,
in a sequential execution,
they are respectively first consumed or produced.
The times are relative to the start of the execution of the relevant conjunct.

The main task of the algorithm is
to divide the execution times of all the conjuncts into chunks
and keep track of when those chunks can execute.
The execution time of \verb|Conj_i|
has one chunk (\verb|Duration_ij|) for each of \verb|Conj_i|'s shared variables
that ends at the time at which that variable is produced or first consumed,
and there is one chunk (\verb|DurationRest_i|) at the end,
during which the call may produce nonshared variables.
Figure~\ref{fig:dep_conj_overlap1} shows that
the production of $A$ divides $p$ into two chunks, ${pA}$ and ${pR}$,
while the consumption of $A$ divides $q$ into ${qA}$ and ${qR}$.

The algorithm processes the chunks in order, and keeps track
of the sequential and parallel execution times of the chunks so far.
When a chunk of \verb|Conj_i| ends with the production of a variable,
we record when that variable is produced,
and the next chunk can start executing immediately.
When a chunk ends with the consumption of a variable,
then in the \emph{sequential} version of \verb|Conj_i|
the next chunk can also execute immediately,
since the values of all the input variables will be available when it starts,
but in the \emph{parallel} version,
the variable may not have been produced yet.
If it has, then \verb|Conj_i| does not need to wait for it;
the left side of figure~\ref{fig:dep_conj_overlap1} shows this case.
However, it is also possible that it has not.
In that case, \verb|Conj_i| will suspend on the variable,
and will resume only when its producer signals that it is available;
the right side of figure~\ref{fig:dep_conj_overlap1} shows this case.
Note that \verb|Var_ij|
will always be in \verb|ProdTimeMap| when we look for it,
because the Mercury mode system reorders conjunctions
to put the producer of each variable before all its consumers.

The version of this algorithm we have actually implemented is 
a bit longer than the one in figure~\ref{fig:dep_par_conj_overlap_middle},
because it also accounts for several forms of overhead:

\begin{itemize}
\item
Creating a spark and adding it to a work queue has a cost.
Every conjunct but the last conjunct incurs this cost
to create the spark for the rest of the conjunction.
\item
It takes some time to take a spark off a spark queue,
create or reuse a context for it, and start its execution.
Every parallel conjunct that is not the first incurs this delay
before it starts running.
\item
The signal and wait operations have a cost.
\item
It takes some time to wake up a context when its wait operation succeeds.
\item
It takes time for each conjunct to synchronize on the barrier
when it has finished its job.
\end{itemize}

\noindent
We can account for every one of these overheads
by adding the estimated cost of the relevant operation to \verb|CurParTime|
at the right point in the algorithm.

In many cases,
the conjunction given to the algorithm shown in figure~\ref{fig:dep_par_conj_overlap_middle}
will contain a recursive call.
In such cases, the speedup computed by the algorithm
reflects the speedup we can expect to get when the recursive call
calls the \emph{original, sequential} version of the predicate.
When the recursive call calls the parallelized version,
we can expect a similar saving (absolute time, not ratio)
on \emph{every} recursive invocation.
How this affects the expected speedup of the top level call
depends on the structure of the recursion.
For the most common recursion structure,
singly recursive predicates like \verb|map_foldl|,
calculating the expected speedup of the top level call is easy,
since we can compute the average depth of recursion
from the relative execution counts of the base and recursive cases.
For some less common structures,
such as doubly recursive predicates like \verb|quicksort|, it is a bit harder,
and for irregular structures in which different execution paths
contain different numbers of recursive calls,
the profiling data gathered by the current version of the Mercury profiler
contains insufficient information to allow our system to determine the
expected speedup.
However, an automated survey of the programs handled by our feedback tool
shows that such predicates are rare;
our system can compute
the expected recursion depth and therefore the expected speedup
for virtually all candidates for parallelization.

So far, we have assumed an unlimited number of CPUs,
which is of course unrealistic.
If the machine has e.g.\ four CPUs,
then the prediction of any speedup higher than four is obviously invalid.
Less obviously,
even a predicted overall speedup of less than four may depend
on more than four conjuncts executing all at once at \emph{some} point.
We have not found this to be a problem yet.
If and when we do,
we intend to extend our algorithm to keep track
of the number of active conjuncts in all active time periods.
Then if a chunk of a conjunct wants to run in a time period
when all CPUs are predicted to be already busy executing previous conjuncts,
we assume that the start of that chunk is delayed until a CPU becomes free.

The limited number of CPUs also means that
there is a limit to how much parallelism we actually \emph{want}.
The spawning off of every conjunct incurs overhead,
but these overheads do not buy us anything if all CPUs are already busy.
That is why our system supports \emph{throttling}.
If a conjunction being parallelized contains a recursive call,
then the compiler can be asked to replace the original sequential conjunction
not with the parallel form of the conjunction,
but with an if-then-else.
The condition of this if-then-else
will test at runtime
whether spawning off a new job is a good idea or not.
If it is, we execute the parallelized conjunction, but
if it is not, we execute the original sequential conjunction.
The condition is obviously a heuristic.
If the heuristic allows the list of runnable jobs to become empty,
then we will not have any work to give to a CPU
that finishes its task and becomes available.
On the other hand,
if the heuristic allows the list of runnable jobs to become too long,
then we incur the overheads of spawning off some jobs unnecessarily.
Currently, on machines with $N$ CPUs,
we prefer to have a total of $M$ running and runnable jobs where $M > N$,
so our heuristic stops spawning attempts
iff the queue already has $M$ entries.
Our current system by default sets $M$ to be $32$ for $N = 4$,
though users can easily override this.

\tr{
\begin{algorithm}
\begin{verbatim}
CurSeqTime_q := 0
CurParTime_q := 0
sort ConsList_q on ConsTime_qi
forall (Var_i, ConsTime_qi) in ConsList_q:
    Duration_qi := ConsTime_qi - CurSeqTime_q
    CurSeqTime_q := CurSeqTime_q + Duration_qi
    ParWantTime_qi := CurParTime_q + Duration_qi
    CurParTime_q := max(ParWantTime_q, Prod_pi)
DurationRest_q := SeqTime_q - CurSeqTime_q
SeqTime_q := CurSeqTime_q + DurationRest_q
ParTime_q := CurParTime_q + DurationRest_q
\end{verbatim}
\caption{Dependent parallel conjunction overlap calculation}
\label{alg:dep_par_conj_overlap_simple}
\end{algorithm}

Algorithm~\ref{alg:dep_par_conj_overlap_simple} shows
a simple version of the algorithm we use to compute ${ParTime}_q$.
Its main input is ${ConsList}_q$,
a list of the variables shared by $p$ and $q$,
together with their times of consumption by $q$.

The main task of the algorithm is to divide up ${SeqTime}_q$ into chunks,
and keep track of when those chunks can execute.
There is one chunk (${Duration}_{qi})$ for each shared variable
that ends at the time at which that variable is first consumed,
and there is one chunk ${DurationRest}$
after the consumption of the last shared variable.
(Figure~\ref{fig:dep_conj_overlap1}
shows the former as ${qA}$ and the latter as ${qR}$.)
The algorithm keeps track of the sequential and parallel execution times of $q$
up to the consumption of the current shared variable.
In the sequential version,
each chunk can execute immediately after the previous chunk,
since the values of the shared variables are all available when $q$ starts.
In the parallel version,
$p$ is producing the shared variables while $q$ is running.
If $p$ has produced the value of ${SV}_i$ by the time $q$ needs it,
there $q$ does not need to wait for it;
the left side of figure~\ref{fig:dep_conj_overlap1} shows this case.
However, it is also possible that $p$ will produce ${SV}_i$
only after the time at which $q$ would like to use it.
In that case, $q$ will suspend on ${SV}_i$,
and will resume only when $p$ signals that it is available;
the right side of figure~\ref{fig:dep_conj_overlap1} shows this case.

On both sides figure~\ref{fig:dep_conj_overlap1}
${SeqTime}_p = 5$ and ${SeqTime}_q = 4$.
On the left side, ${ConsTime}_{qA} = 2$,
and therefore ${Duration}_{qA} = 2$ and ${DurationRest}_{qA} = 2$,
Since ${ProdTime}_{pA} = 1$,
the first update of ${CurParTime}_q$ sets it to $2$,
and the second sets it to $4$, so ${ParTime}_q = 4$.
On the right side, ${ConsTime}_{qA} = 1$,
and therefore ${Duration}_{qA} = 1$ and ${DurationRest}_{qA} = 3$,
Since ${ProdTime}_{pA} = 4$,
the first update of ${CurParTime}_q$ sets it to ${max}(1, 4) = 4$,
and the second sets it to $4+3 = 7$, so ${ParTime}_q$ is 7.

\begin{table}
\begin{center}
\begin{tabular}{l|rr}
 & \multicolumn{1}{|c}{Cost}
 & \multicolumn{1}{|c}{Local use of \code{Acc1}} \\
\hline
\code{M}  &   1,625,050 & none \\
\code{F}  &           3 & ${Prod}_{Acc1}$ =         3 \\
\mapfoldl &   1,625,054 & ${Cons}_{Acc1}$ = 1,625,051 \\
\end{tabular}
\end{center}
\caption{Rounded profiling data for \mapfoldl}
\label{tab:prof_data_map_foldl}
\end{table}

\begin{figure}[tb]
\begin{verbatim}
map_foldl_par(_, _, [], Acc, Acc).
map_foldl_par(M, F, [X | Xs], Acc0, Acc) :-
    (
        M(X, Y),
        F(Y, Acc0, Acc1)
    ) &
    map_foldl_par(M, Xs, Acc1, Acc).
\end{verbatim}
\caption{Parallel \mapfoldl}
\label{fig:map_foldl_par}
\end{figure}

To see how the algorithm works on realistic data,
consider the \mapfoldl example in figure~\ref{fig:map_foldl}.
Table \ref{tab:prof_data_map_foldl} gives
the approximate costs of the calls in the recursive clause of \mapfoldl
when used in a Mandelbrot image generator.
Each call to $M$ draws a row,
while $F$ appends the new row
onto the list of the rows already drawn.
The table also shows when $F$ produces ${Acc1}$
and when the recursive call consumes ${Acc1}$.
The costs were collected from a real execution using Mercury's deep profiler
and then rounded to make mental arithmetic easier.

Figure~\ref{fig:map_foldl_par} shows the best parallelization of
\mapfoldl.
When evaluating the speedup for this parallelization,
${Cons}_{{F} {Acc1}} = 1,625,000 + 2 = 1,625,002$, and
${Cons}_{{map\_foldl} {Acc1}} = 1,625,000 + 2 + 1,620,000 = 3,245,002$.
\zoltan{Show the rest of the algorithm's execution when this question is answered.}
\pbone{I would have fixed this but I don't know what these formulas mean,
Maybe I'm confused by notation}

The speedup computed by algorithm~\ref{alg:dep_par_conj_overlap_simple}
applies only when the recursive call
calls the original sequential version of the predicate.
When the recursive call calls the parallelized version,
the maximum speedup available (assuming an unlimited number of CPUs)
depends on the structure of the recursion.
The profiling data gives us \tr{$E$,}
the number of entry calls to the procedure from the higher clique,
and \tr{$R_i$,} the number of recursive calls at each recursive call site.
From these, and the structure of the procedure's code,
we can calculate the average depth of recursion in most cases.

For singly recursive predicates like \mapfoldl,
there is only one recursive call site,
and the depth of recursion is simply $R_1/E$.
For example, if $E = 2$ and $R_1 = 20$,
then the average call sequence to the procedure
has one entry call followed by the recursive calls.
It also means that the average call sequence
has ten calls that cause the procedure to execute its recursive clause,
the clause containing the conjunction being parallelized,
followed by one call that executes the base clause.
From this, we can deduce that if ${SeqSaving} = {SeqTime} - {ParTime}$
is the time saving we get from parallelizing the top conjunction
if the recursive call calls the original sequential version of the procedure,
then making the recursive call call the parallelized version of the procedure
would yield a time saving of ${ParSaving} = {SeqSaving} * R_1/E$
if we have enough CPUs to execute all the $R_1/E$ iterations in parallel.
\peter{I think this needs more explanation.  It doesn't look right to me.
I would expect it to be ${ParSaving} = {SeqTime} - {ParTime} * R_1/E$.}
This assumes that we get the same savings at each iteration,
\peter{I find this sentence weakens the claim, rather than strengthing it.
Can we just add ``, assuming we get the same savings at each iteration'' at
the end of the previous sentence?}
but this is reasonable,
since what we are doing is essentially executing
the different iterations of a loop in parallel,
and we have no reason to believe that the savings
from executing iteration $k$ in parallel with iteration $k+1$
would vary systematically based on the value of $k$.

Some singly recursive predicates have more than one recursive clause,
each with one recursive call site.
Suppose there are $n$ call sites, with execution counts $R_1 \ldots R_n$.
The overall time saving from parallelizing the conjunct
that contains the call site associated $R_i$
has to be multiplied by the fraction of recursive calls
that execute the conjunction being parallelized:
${ParSaving} = {SeqSaving} * R_i/E * R_i/\sum_{j=1}^n R_j$.

For doubly recursive predicates like \code{quicksort}, $R_1 = R_2$,
and just under half of their calls invoke the recursive clause,
so for them, ${ParSaving} = {SeqSaving} * R_1/(2 * E)$.
\zoltan{check the math}
\peter{I think it's a bit confusing to use $R_1$ and $R_2$ to name the
recursive calls to quicksort, when above they've named the sole recursive
call from different clauses.  Maybe use $R^i_j$ to indicate the $i^{th}$
recursive call from the $j^{th}$ clause.  Then you can use $R^1_i$ in the
previous paragraph, and $R_1^1$ and $R_1^2$ in this one.}

All these calculations show that the available parallelism
can be greater than the number of CPUs.
If the machine has e.g.\ four CPUs,
then we do not actually want to spawn off
hundreds of iterations for parallel execution,
since parallel execution actually has several forms of overhead:

\begin{description}
\item[SparkCost]
is the cost of creating a spark and adding it to the local spark stack.
In a parallel conjunction,
every conjunct that is not the last conjunct incurs this cost
to create the spark for the rest of the conjunction.
\item[SparkDelay]
is the estimated length of time between the creation of a spark
and the beginning of its execution on another engine.
Every parallel conjunct that is not the first incurs this delay
before it starts running.
\item[SignalCost]
is the cost of signaling a future.
\item[WaitCost]
is the cost of waiting on a future.
\item[ContextWakeupDelay]
is the estimated time that it takes for a context to resume execution
after being placed on the runnable queue,
assuming that the queue is empty and there is an idle engine.
\item[BarrierCost]
is the cost of executing the operation
that synchronizes all the conjuncts at the barrier
at the end of the conjunction.
\end{description}

Because of these overheads, our system uses \emph{throttling}.
If a conjunction being parallelized contains a recursive call,
then the compiler will replace the original sequential conjunction
not with the parallel form of the conjunction,
but with an if-then-else.
The condition of this if-then-else
will test at runtime
whether spawning off a new job is a good idea or not.
If it is, we execute the parallelized conjunction,
if it is not, we execute the original sequential conjunction.
The condition is obviously a heuristic.
If the heuristic allows the list of runnable jobs to become empty,
then we will not have any work to give to a CPU
that finishes its task and becomes available.
On the other hand,
if the heuristic allows the list of runnable jobs to become too long,
then we incur the overheads of spawning off some jobs unnecessarily.
Currently, on machines with $N$ CPUs,
we prefer to have a total of $M$ running and runnable jobs where $M > N$,
so our heuristic stops spawning attempts
iff the queue already has $M$ entries.
Our current system by default sets $M$ to be $32$ for $N = 4$,
though users can easily override this.

The overheads of parallel execution can also affect conjunctions
that do not contain recursive calls:
a conjunction that looks worth parallelizing if you ignore overheads
may look not worth parallelizing if you take them into account.
This is why our system actually uses
algorithm~\ref{alg:dep_par_conj_overlap_complete},
a version of algorithm~\ref{alg:dep_par_conj_overlap_simple}
that accounts for overheads.

\begin{algorithm}
\begin{verbatim}
find_par_time(Conjs) returns TotalParTime:
N := length(Conjs)
ProdTimeMap := empty
FirstConjTime := 0
TotalParTime  := 0
for i in 1 to N:
  CurSeqTime := 0
  CurParTime := (SparkCost + SparkDelay) * (i-1)
  if i != N:
    CurParTime := CurParTime + SparkCost
  sort ProdConsList_i on Time_ij
  forall (Var_ij, Time_ij) in ProdConsList_i:
    Duration_ij := Time_ij - CurSeqTime
    CurSeqTime := CurSeqTime + Duration_ij
    if Conj_i produces Var_ij:
      CurParTime := CurParTime + Duration_ij + SignalCost
      ProdTimeMap[Var_ij] := CurParTime
    else Conj_i must consume Var_ij:
      ParWantTime := CurParTime + Duration_ij
      CurParTime := max(ParWantTime, ProdTimeMap[Var]) + WaitCost
      if ParWantTime < ProdTimeMap[Var_ij]:
        CurParTime := CurParTime + ContextWakeupDelay
  DurationRest := SeqTime_i - CurSeqTime
  CurParTime := CurParTime + DurationRest + BarrierCost
  if i == 1:
    FirstConjTime = CurParTime
  TotalParTime := max(TotalParTime, CurParTime)
if TotalParTime > FirstConjTime:
  TotalParTime := TotalParTime + ContextWakeupDelay
\end{verbatim}
\caption{Dependent parallel conjunction complete algorithm}
\label{alg:dep_par_conj_overlap_complete}
\end{algorithm}

Algorithm~\ref{alg:dep_par_conj_overlap_complete}
can also handle $n$-way conjunctions for $n>2$.
Since the Mercury mode system reorders conjunctions
to ensure that data flows only to the right,
in a two-conjunct conjunction,
the left conjunct can only produce
the variables it shares with the right conjunct
and the right conjunct can only consume
those variables.
However, in longer conjunctions,
the conjuncts in the middle
can both consume variables produced by conjuncts on their left
and produce variables consumed by conjuncts on their right.
This is why our algorithm associates with \code{Conj\_i}, the $i$th conjunct,
\code{ProdConsList\_i}:
the list of shared variables that \code{Conj\_i} either produces or consumes,
together with their times of production and first consumption respectively.
This is a generalization of \code{ConsList\_q} in
algorithm~\ref{alg:dep_par_conj_overlap_simple}.
We also need to generalize \code{Prod\_pi},
because the time at which a non-first conjunct produces a variable
can and usually will be affected
by the overheads and/or synchronization delays suffered by that conjunct.
This is why we use \code{ProdTimeMap},
which maps each shared variable to its time of production.

The main body of the algorithm consists of two nested loops.
The outer loop loops over all the conjuncts from left to right,
because the execution of a conjunct can be affected
by the conjuncts to its left
(through the time at which they produce the shared variables it consumes),
but not by the conjuncts to its right.
The first few lines of the outer loop body
(the first two assignments to \code{CurParTime})
compute for each conjunct
the time at which that conjunct can start execution.

The inner loop loops over all the components of the current conjunct,
such as \code{pA} and \code{pR}
from figure~\ref{fig:dep_conj_overlap1}.
Just as our previous algorithm did,
this loop updates the simulated current time
in both the original sequential execution of the conjunct (\code{CurSeqTime})
and in its modified parallelized execution (\code{CurParTime}).
For time components that end in the consumption of a variable,
we do what we did before,
but also reflect the cost of the wait operation needed for the consumption.
For time components that end in the production of a variable,
we record the time at which
that variable would be available in the parallel execution;
this will be when the producer finishes executing the signal operation on it.

We use \code{TotalParTime} to keep track of the ending time
of the parallel conjunct that ends last.
We also remember, in \code{FirstConjTime},
the time at which the first conjunct finishes.
The reason we do this is because
our runtime system requires that
when the parallel conjunction finishes,
execution must continue in the context
that entered the parallel conjunction in the first place.
In our implementation, this context will execute the first conjunct.
If the last conjunct to finish is the first conjunct,
it can continue on without delay;
if the last conjunct to finish is some other conjunct,
then we need to free its context,
and switch to executing the original context,
which became idle when the first conjunct finished.
The last two lines reflect this cost.

}

\section{Choosing how to parallelize a conjunction}
\label{sec:howto}

A conjunction with $n > 2$ conjuncts
can be converted into several different parallel conjunctions.
Converting all the commas into ampersands
(e.g.\ \code{c1, c2, c3} into \code{c1 \& c2 \& c3})
yields the most parallelism.
Unfortunately, this will often be \emph{too} much parallelism,
because in practice many conjuncts are unifications
and arithmetic operations whose execution takes very few instructions.
Executing such conjuncts in their own threads
costs far more in overheads than they save by running in parallel.
Therefore in most cases,
we want to create parallel conjunctions with $k < n$ conjuncts,
each consisting of a contiguous sequence
of one or more of the original sequential conjuncts,
effectively partitioning the original conjuncts into groups.

\begin{figure}
\figrule
\begin{center}
\begin{verbatim}
global NumEvals := 0
find_best_partition(InitPartition, InitTime, LaterConjs)
    returns <FinalTime, FinalPartitionSet>:
  switch on LaterConjs:
  when LaterConjs = []:
    return <InitTime, {InitPartition}>
  when LaterConjs = [Head | Tail]:
    Extend := all_but_last(InitPartition) ++ [last(InitPartition) ++ [Head]]
    AddNew := InitPartition ++ [Head]
    ExtendTime := find_par_time(Extend)
    AddNewTime := find_par_time(AddNew)
    NumEvals := NumEvals + 2
    if ExtendTime < AddNewTime:
      BestExtendSoln := find_best_partition(Extend, ExtendTime, Tail)
      let BestExtendSoln be <BextExTime, BestExPartSet>
      if NumEvals < PreferLinearEvals:
        BestAddNewSoln := find_best_partition(AddNew, AddNewTime, Tail)
        let BestAddNewSoln be <BestANTime, BestANPartSet>
        if BestExTime < BestANTime:
          return BestExtendSoln
        else if BestExTime = BestANTime:
          return <BextExTime, BestExPartSet union BestANPartSet>
        else:
          return BestAddNewSoln
      else:
        return BestExtendSoln
    else:
      <symmetric with the then case>
\end{verbatim}
\end{center}
\caption{Search for the best parallelization}
\label{fig:best_par_search}
\figrule
\vspace{-2\baselineskip}
\end{figure}

For any conjunction to be worth parallelizing,
it should contain two or more expensive goals.
Our main algorithm (figure \ref{fig:best_par_search} works on the list
of conjuncts
from the first expensive goal to the last.
This will be the middle of original conjunction,
with (possibly empty) lists of cheap goals before it and after it.
Our initial search assumes that
the set of conjuncts in the parallel conjunction we want to create
is exactly the set of conjuncts in the middle.
A post-processing step then removes that assumption.

If the middle sequence has $n$ conjuncts,
then there are $n-1$ AND operations between them,
each of which can be either sequential or parallel.
There are then $2^{n-1}$ combinations,
all but one of which are parallelizations.
That is a large space to search for the \emph{best} parallelization,
and it would be larger still if we allowed code reordering,
that is, parallel conjuncts consisting of
a \emph{non}contiguous sequence of the original conjuncts.
We explore this space with a search algorithm,
\code{find\-\_\-best\-\_\-par\-ti\-tion}, which
we invoke with the empty list as \code{InitPartition},
zero as \code{InitTime}, and the list of middle conjuncts as \code{LaterConj}.
\code{InitPartition} expresses a partition of an initial sequence
of the middle goals into parallel conjuncts
whose estimated execution time is \code{InitTime},
and considers whether it is better to add the next middle goal
to the last existing parallel conjunct (\code{Extend}),
or to put it into a new parallel conjunct (\code{AddNew}).
It explores extensions of the better of the resulting partitions first.
If the search is still under the limit on the number of evaluations,
it explores the worse partition as well,
which is an exponential search.
When it hits the limit, 
it switches to a linear search;
we explore the more promising partition first
to make this search more effective.
(This limit ensures that the algorithm runs in reasonable time.)
The algorithm returns a set of equal best parallelizations so far,
``best'' being measured by
\iclp{a version of the algorithm in figure~\ref{fig:dep_par_conj_overlap_middle} that
computes the estimated parallel execution time \emph{including} overheads.}
\tr{algorithm \ref{alg:dep_par_conj_overlap_complete},
that is, the estimated parallel execution time including overheads.}

There are some simple ways to improve this algorithm.
\vspace{-2mm}
\begin{itemize}
\item
Most invocations of \verb|find_par_time| specify a partition
that is an extension of a partition processed in the recent past.
In such cases, \verb|find_part_time| should do its task
incrementally, not from scratch.
\item
If the expected execution time
for the candidate partition currently being considered
is already greater than the fastest existing complete partition,
we can stop exploring that branch;
it cannot lead to a better solution.
\tr{
(This is the idea of branch-and-bound algorithms.)
}
\item
Sometimes consecutive conjuncts do things that are
obviously a bad idea to do in parallel, such as building a ground term.
The algorithm should treat these as a single conjunct.
\tr{
\item
graph of dependencies
\item
take total CPU utilization into account,
at least by using it to break ties on overall CPU time
}
\end{itemize}
\vspace{-2mm}

\noindent
At the completion of the search,
we select one of the equal best parallelizations,
and post-process it to adjust both edges.
Suppose the best parallel form of the middle goals is $P_1~\&~\ldots~\&~P_p$,
where each $P_i$ is a sequential conjunction.
We compare the execution time of $P_1~\&~\ldots~\&~P_p$
with that of $P_1,~(P_2~\&~\ldots~\&~P_p)$.
If the former is slower,
which can happen if $P_1$ produces its outputs at its very end
and the other $P_i$ consume those outputs at their start,
then we conceptually move $P_1$ out of the parallel conjunction
(from the ``middle'' part of the conjunction to the ``before'' part).
We keep doing this for $P_2$, $P_3$ etc until either
we find a goal worth keeping in the parallel conjunction,
or we run out of conjuncts.
We also do the same thing at the other end of the middle part.
This process can shrink the middle part.

In cases where we do not shrink an edge, we can consider expanding that edge.
Normally, we want to keep cheap goals out of parallel conjunctions,
since more conjuncts tends to mean
more shared variables and thus more synchronization overhead,
but sometimes this consideration is overruled by others.
Suppose the goals before the conjuncts in $P_1~\&~\ldots~\&~P_p$
in the original conjunction were $B_1,~\ldots,~B_b$
and the goals after it $A_1,~\ldots,~A_a$,
and consider $A_1$ after $P_p$.
If $P_p$ finishes before the other parallel conjuncts,
then executing $A_1$ just after $P_p$ in $P_p$'s context
may be effectively free:
the last context could still arrive at the barrier at the same time,
but this way, $A_1$ would have been done by then.
Now consider $B_b$ before $P_1$.
If $P_1$ finishes before the other parallel conjuncts,
\emph{and} if none of the other conjuncts wait for variables produced by $P_1$,
then executing $B_b$ in the same context as $P_1$ can be similarly free.

We loop from $i=b$ down towards $i=1$, and check whether
including $B_i,~\ldots,~B_b$ at the start of $P_1$ is improvement.
If not, we stop; if it is, we keep going.
We do the same from the other end.
The stopping points of the loops of the contraction and expansion phases
dictate our preferred parallel form of the conjunction, which
(if we shrunk the middle at the left edge and expanded it at the right)
will look something like
$B_1,$ $\ldots,$ $B_{b},$ $P_1,$ $\ldots~P_k,$
$(P_{k+1}$ $\&$ $\ldots$ $\&$ $P_{p-1}$ $\&$ $(P_p,$ $A_1,$ $\ldots,$ $A_j)),
A_{j+1},$ $\ldots,$ $A_a$.
If this preferred parallelization is better than
the original sequential version of the conjunction by at least 1%
then we include a recommendation for its conversion to this form
in the feedback file we create for the compiler.

\tr{
\section{Pragmatic issues}
\label{sec:pragmatic}

\emph{Dynamic context}
The algorithms in sections~\ref{sec:overlap} and~\ref{sec:howto}
work on profiling data that shows the behavior of a procedure
in the context given by a particular chain of ancestors.
Many procedures are of course called from multiple ancestor contexts.
What happens when our analysis of the behavior of the same procedure
yields different results for different ancestor contexts?

At the moment, for any procedure
that our analysis indicates is worth parallelizing in any context,
we pick one particular parallelization (usually there is only one anyway),
and transform the procedure accordingly.
This gets the benefit of parallelization when it is worthwhile,
but incurs its costs even in contexts when it is not.
In the future, we plan to fix this using multi-version specialization.
For every procedure with different parallelization recommendations,
we intend to create a specialized version for each recommendation,
leaving the original sequential version.
This will of course require the creation of specialized versions
of its parent, grandparent etc procedures,
until we get to an ancestor procedure
which occurs in the common prefix of all the conflicting ancestor contexts.

\emph{Parallelizing children vs ancestors}
What happens when we decide that a conjunction that should be parallelized
has an ancestor that we decided should also be parallelized?
We can
(1) parallelize only the ancestor,
(2) parallelize only this conjunction, or
(3) parallelize both

\zoltan{Does the implementation actually do this now?}
We choose among the other three alternatives
by evaluating the speedup you get from each of them, and just pick the best.
This reevaluation must take into account
the fact that for each invocation of the ancestor conjunction,
we will invoke the current conjunction many times,
and that therefore we will incur both the overheads and the benefits
of parallelizing the current conjunction many times.
The profile will give the actual number.

\emph{Parallelizing branched goals}
Many programs have code that looks like this:
\begin{verbatim}
( if ... then
    ... expensive call 1 ...
else
    ... cheap goal ...
),
expensive call 2
\end{verbatim}
If the condition of the if-then-else succeeds only rarely,
then the average cost of the if-then-else
may be below the threshold of what we consider to be an expensive goal.
We therefore would not even consider
parallelizing the top-level conjunction,
rightly considering that its overheads would probably outweight its benefits.

What we want to do in such cases
is execute just the two expensive calls in parallel,
which would be equivalent to parallelizing the conjunction
in the then part of this transformed goal:
\begin{verbatim}
( if ... then
    ... expensive call 1 ...
    expensive call 2
else
    ... cheap goal ...
    expensive call 2
)
\end{verbatim}
We intend to change our feedback tool to detect such situations,
and if found, to recommend
some equivalence-preserving transformations for the compiler to apply
before parallelizing some of the resulting conjunctions.

\emph{Garbage collector issues}
The Mercury implementation uses the Boehm-Demers-Weiser
conservative collector for C \zoltan{add cite} to manage memory.
This system has worse overheads
for parallel programs than for sequential programs.
First, even though this collector uses
a separate memory pool for each mutator thread
(and hence, in our system, for each Mercury engine),
you still need synchronization to access the global pool
when a local pool runs out.
Second, this collector
does not support incremental collection for parallel programs,
and a full collection stops all threads,
and thrashes the caches of their CPUs.
We therefore ran our benchmarks with the collector tuned
to use large local pool sizes
and to grow the size of the global pool more quickly than usual.
These settings significantly improved
the performance of the sequential programs as well.

}

\section{Performance results}
\label{sec:perf}

We tested our system on three benchmark programs:
matrix multiplication, a mandelbrot image generator and a raytracer.
Matrixmult has abundant independent AND-parallelism.
Mandelbrot uses the actual \code{map\_foldl} predicate
from figure~\ref{fig:map_foldl}
to iterate over rows of pixels.
Raytracer does not use \code{map\_foldl},
but does use a similar code structure to perform a similar task.
This is not an accident:
\emph{many} predicates use this kind of code structure,
partly because programmers in declarative languages
often use accumulators to make their loops tail recursive.

We ran all three programs
with one set of input parameters to collect profiling data,
and with a \emph{different} set of input parameters to produce
the timing results in the following table.
All tests were run on
a Dell Optiplex 755 PC with a 2.4~GHz Intel Core 2 Quad Q6600 CPU
running Linux 2.6.31.
Each test was run ten times;
we discarded the highest and lowest times, and averaged the rest.

\begin{table}[tb]
\begin{center}
\begin{tabular}{llrrrrr}
\hline \hline
\multicolumn{1}{c}{\textbf{Program}} &
\multicolumn{1}{c}{\textbf{Par}}    & 
\multicolumn{1}{c}{\textbf{1 CPU}}   & 
\multicolumn{1}{c}{\textbf{2 CPUs}}  & 
\multicolumn{1}{c}{\textbf{3 CPUs}}  & 
\multicolumn{1}{c}{\textbf{4 CPUs}}  \\
\hline
matrixmult & indep    & 14.6 (0.75) &  7.5 (1.47) &  7.0 (1.66) &  5.2 (2.12) \\
seq 11.0   & naive    & 14.6 (0.75) &  7.6 (1.45) &  5.2 (2.12) &  5.2 (2.12) \\
par 14.6   & overlap  & 14.6 (0.75) &  7.5 (1.47) &  6.2 (1.83) &  5.2 (2.12) \\
\hline
mandelbrot & indep    & 35.2 (0.95) & 35.1 (0.95) & 35.2 (0.95) & 35.3 (0.95) \\
seq 33.4   & naive    & 35.4 (0.94) & 18.0 (1.86) & 12.1 (2.76) &  9.1 (3.67) \\
par 35.2   & overlap  & 35.6 (0.94) & 17.9 (1.87) & 12.1 (2.76) &  9.1 (3.67) \\
\hline
raytracer  & indep    & 26.2 (0.87) & 26.3 (0.86) & 26.1 (0.87) & 26.2 (0.87) \\
seq 22.7   & naive    & 25.3 (0.90) & 16.0 (1.42) & 11.2 (2.03) &  9.4 (2.42) \\
par 26.5   & overlap  & 25.1 (0.90) & 16.0 (1.42) & 11.2 (2.03) &  9.4 (2.42) \\
\hline \hline
\end{tabular}
\end{center}
\vspace{-2\baselineskip}
\end{table}

Each group of three rows reports the results for one benchmark.
The first column shows the benchmark name,
the runtime of the program when compiled for sequential execution, and
its runtime when compiled for parallel execution
but without enabling auto-parallelization.
This shows the overhead of support for parallel execution
when it does not buy any benefits.
We auto-parallelized each program three different ways:
executing expensive goals in parallel
only when they are independent (``indep'');
even if they are dependent, regardless of overlap (``naive'');  and
even if they are dependent, but only if they have good overlap (``overlap'').
The last four columns give the runtime in seconds
of each of these versions of the program
on 1, 2, 3 and 4 CPUs,
with speedups compared to the sequential version.

The parallel version of the Mercury system
needs to use a real machine register
to point to thread-specific data,
such as each engine's abstract machine registers.
On x86s, this leaves only one real register for the Mercury abstract machine,
so compiling for parallelism but not using it
yields a slowdown ranging from 5\% on mandelbrot to 25\% on matrixmult.
(We observe such slowdowns for other programs as well.)
On one CPU, autoparallelization gets only this slowdown,
plus the (small) additional overheads of all the parallel conjunctions
that cannot get any parallelism.

The parallelism in the main predicate of matrixmult is independent,
Overlap parallelizes the program the same way as indep,
so it gets the same speedup.
The numbers look different for 3 CPUs,
but all the runs for both versions actually took either 5.2 or 7.5 seconds,
depending (we think) on which way
the OS arranged the engines across the two CPU die of the Q6600;
the indep version just happened to get the 7.5s arrangement fewer times.
For naive, all the runs just happened to take 5.2 seconds,
even though naive creates a worse parallelization than either indep or overlap:
during the expansion phase we described in section~\ref{sec:howto},
it includes an extra goal in the first of the parallel conjuncts;
this makes the conjunction dependent, which adds some overhead.
Naive also executes the code that does the matrix multiplication
in parallel with the goals that create its inputs,
which also adds overhead without speedup.
These overheads are too small to affect the results.

In mandelbrot and raytracer, all the parallelism is dependent,
which is why indep gets no speedup for them.
For mandelbrot, naive and overlap get speedups
that are as good as one can reasonably expect:
$35.2/9.1 = 3.87$ on four CPUs over the one CPU case.
For matrixmult and raytracer, the speedups they get,
2.12 and 2.42 on four CPUs,
also turn out to be pretty good when one takes a closer look.

For matrixmult, the bottleneck is almost certainly CPU-memory bandwidth.
Each step in this program does only one multiply and one add (both integer)
before creating a new cell on the heap and filling it in.
On current CPUs, the arithmetic takes much less time than the memory writes,
and since the new cells are never accessed again, caches do not help,
which makes it easy to saturate the memory bus, even when using only three CPUs.

The raytracer is very memory-allocation-intensive,
because it does lots of FP arithmetic,
and the Mercury backend we are using always boxes floating point numbers,
so each floating point operation requires
the creation of a new cell on the heap.
Because of this, memory bandwidth may also be an issue for it,
but its bigger problem is GC;
while GC takes only about 5\% of the runtime when run on one CPU,
it takes almost 40\% of the runtime when run on four CPUs,
even though we used four marker threads.
(For fairness, we used four marker threads
regardless of how many CPUs the Mercury code used.)
Given this fact, the best speedup we can hope for is
$(4 \times 0.6 + 0.4)/(0.6 + 0.4) = 2.8$,
and we do come pretty close to that.

GC becomes more expensive with more CPUs
not only because of increased contention,
but also because the GC has more work to do:
with more contexts being spawned, there are more stacks for it to scan.
We have tested versions of the raytracer in which
each spawned-off goal computed the pixels for several rows, not just one,
and these versions yield speedups of about 3.3 on four CPUs.
These versions spawn many fewer contexts, thus putting much less load
on the GC.
This shows that
program transformations that cause more work to be done in each context
are likely to be a promising area for future work.

Most small programs like these benchmarks
have only one loop that dominates their runtime.
In all three of these benchmarks, and in many others,
the naive and overlap methods will parallelize the same loops,
and usually the same way;
they tend to differ only in how they parallelize code
that executes much less often (typically only once)
whose effect is lost in the noise.
The raw timings show a great deal of variability:
we have seen two consecutive runs of the same program on the same data
differ in their runtime by as much as 15\%.
Some of this variability remains even after filtering and averaging.

To see the difference between naive and overlap,
we need to look at larger programs.
Our standard large test program is the Mercury compiler, which contains
53 conjunctions with two or more expensive goals.
Of these, 52 are dependent,
and only 31 have an overlap
that leads to a predicted local speedup of more than 1\%,
our default threshold.
Our algorithms can thus prevent
the unproductive parallelization of $53-31=22$ of these conjunctions.
Unfortunately, programs that are large and complex enough
to show a performance effect from this saving
also tend to have large components
that cannot be profitably parallelized with existing techniques,
which means that (due to Amdahl's law)
our autoparallelization system cannot yield overall speedups for them yet.

On the bright side,
our feedback tool generates feedback files
in less than a second from the profiles of small programs like these benchmarks,
and in only a minute or two even from much larger profiles.
The extra time taken by the Mercury compiler
when it follows the recommendations in feedback files
is so small that it is not noticeable.

\section{Related work and conclusion}
\label{sec:conc}

Mercury's strong mode and determinism systems
greatly simplify the parallel execution of logic programs.
The information gathered by semantic analysis in Mercury
makes it easy to solve most of the problems faced by the
designers of parallel versions of Prolog and Prolog-like languages.
These include testing the independence of goals
in systems that support only independent AND-parallelism
and discovering producer-consumer relationships
in systems that also support dependent AND-parallelism,
such as \citeN{DBLP:journals/tcs/GrasH09}.
They also make it possible to \emph{avoid} having to solve some tough problems,
the main example being how to execute nondeterministic conjuncts in parallel
without excessive overhead.

Most research in parallel logic programming so far
has focused on trying to solve these problems
of getting parallel execution to \emph{work} well,
with only a small fraction trying to find
when parallel execution would actually be \emph{worthwhile}.
Almost all previous work on automatic parallelization 
has focused on granularity control:
parallelizing only computations that are expensive enough
to make parallel execution
worthwhile \cite{harris_07_feedback_imp_par,lopez96:distance_granularity},
and properly accounting for the overheads
of parallelism itself \cite{shen_98_granularity-control}.
Most of the rest has tried to find opportunities
to exploit independent AND-parallelism
during the execution of otherwise-dependent conjunctions
\cite{DBLP:journals/jlp/MuthukumarBBH99,DBLP:conf/lopstr/CasasCH07}.

Our experience with our feedback tool shows that
for Mercury programs, this is far from enough.
For most programs,
it finds enough conjunctions with two or more expensive conjuncts,
but almost all are dependent,
and, as we mention in section~\ref{sec:perf},
many of these have too little overlap to be worth parallelizing.

We know of only three attempts to estimate the overlap
between parallel computations.
One was in the context of speculative execution in imperative programs.
Given two successive blocks of instructions,
\cite{von_Praun:2007:implicit_parallelism_with_ordered_transactions}
decides whether the second block should be executed speculatively
based on the difference between the addresses of two instructions,
one that writes a value to a register and one that reads from that register.
This works if instructions take a bounded time to execute,
but in the presence of call instructions
this heuristic will not be at all accurate.

Another attempt was a previous auto-parallelization project for
Mercury \cite{tannier:2007:parallel-mercury}.
This used the number of shared variables between conjuncts
as a measure of the dependency between goals,
and as a predictor of the likely overlap.
While two conjuncts are indeed less likely
to have useful parallel overlap if they have more shared variables,
we have found this heuristic too inaccurate to be useful.

The most closely related work to ours
generated parallelism annotations for the ACE and/or-parallel system
\cite{Pontelli97automaticcompile-time}.
This system used, much as we do,
estimates of the costs of calls
and of the times at which variables are produced and consumed.
However, it produced its estimates through static analysis of the program.
This can work for small programs,
where the call trees of the relevant calls can be quite small and regular.
In large programs, the call trees of the expensive calls
are almost certain to be both tall and wide,
with a huge gulf between best-case and worst-case behavior.
Using profiling data is the only way
for an automatic parallelization system to find out
what the \emph{typical} behavior of such calls is.

Our system's predictions of the likely speedup from parallelizing a conjunction
are also fallible, since they currently ignore several relevant issues,
including cache effects
and the effects of bottlenecks
such as CPU-memory buses and stop-the-world garbage collection.
However, our system seems to be a sound basis for such further refinements.
In the future, we plan to support parallelization as a specialization:
applying a specific parallelization only when a predicate is called
from a specific parent, grandparent or other ancestor.
We also plan to modify our feedback tool
to accept several profiling data files,
with a priority scheme to resolve any conflicts.
We thank the rest of the Mercury team,
and Tom Conway and Peter Wang in particular,
for creating the infrastructure we build upon,
and the anonymous referees for their suggestions.

\bibliographystyle{acmtrans}
\bibliography{dep_par_conj_overlap}

\begin{thebibliography}{}

\bibitem[\protect\citeauthoryear{Bevemyr, Lindgren, and Millroth}{Bevemyr
  et~al\mbox{.}}{1993}]{bevemyr:reform}
{\sc Bevemyr, J.}, {\sc Lindgren, T.}, {\sc and} {\sc Millroth, H.} 1993.
\newblock Reform {P}rolog: the language and its implementation.
\newblock In {\em Proceedings of the Tenth International Conference on Logic
  Programming}. Budapest, Hungary, 283--298.

\bibitem[\protect\citeauthoryear{Casas, Carro, and Hermenegildo}{Casas
  et~al\mbox{.}}{2007}]{DBLP:conf/lopstr/CasasCH07}
{\sc Casas, A.}, {\sc Carro, M.}, {\sc and} {\sc Hermenegildo, M.~V.} 2007.
\newblock Annotation algorithms for unrestricted independent {AND}-parallelism
  in logic programs.
\newblock In {\em Proceedings of the 17th International Symposium on
  Logic-based Program Synthesis and Transformation}. Lyngby, Denmark, 138--153.

\bibitem[\protect\citeauthoryear{Gras and Hermenegildo}{Gras and
  Hermenegildo}{2009}]{DBLP:journals/tcs/GrasH09}
{\sc Gras, D.~C.} {\sc and} {\sc Hermenegildo, M.~V.} 2009.
\newblock Non-strict independence-based program parallelization using sharing
  and freeness information.
\newblock {\em Theoretical Computer Science\/}~{\em 410,\/}~46, 4704--4723.

\bibitem[\protect\citeauthoryear{Halstead}{Halstead}{1984}]{multilisp}
{\sc Halstead, R.~H.} 1984.
\newblock Implementation of {MultiLisp}: {L}isp on a multiprocessor.
\newblock In {\em Proceedings of the 1984 ACM Symposium on List and Functional
  Programming}. Austin, Texas, 9--17.

\bibitem[\protect\citeauthoryear{Harris and Singh}{Harris and
  Singh}{2007}]{harris_07_feedback_imp_par}
{\sc Harris, T.} {\sc and} {\sc Singh, S.} 2007.
\newblock Feedback directed implicit parallelism.
\newblock {\em SIGPLAN Notices\/}~{\em 42,\/}~9, 251--264.

\bibitem[\protect\citeauthoryear{Lopez, Hermenegildo, and Debray}{Lopez
  et~al\mbox{.}}{1996}]{lopez96:distance_granularity}
{\sc Lopez, P.}, {\sc Hermenegildo, M.}, {\sc and} {\sc Debray, S.} 1996.
\newblock A methodology for granularity-based control of parallelism in logic
  programs.
\newblock {\em Journal of Symbolic Computation\/}~{\em 22,\/}~4, 715--734.

\bibitem[\protect\citeauthoryear{Marlow, Jones, and Singh}{Marlow
  et~al\mbox{.}}{2009}]{simonmar_2009_multicore_rts}
{\sc Marlow, S.}, {\sc Jones, S.~P.}, {\sc and} {\sc Singh, S.} 2009.
\newblock Runtime support for multicore {H}askell.
\newblock {\em SIGPLAN Notices\/}~{\em 44,\/}~9, 65--78.

\bibitem[\protect\citeauthoryear{Muthukumar, Bueno, de~la Banda, and
  Hermenegildo}{Muthukumar
  et~al\mbox{.}}{1999}]{DBLP:journals/jlp/MuthukumarBBH99}
{\sc Muthukumar, K.}, {\sc Bueno, F.}, {\sc de~la Banda, M. J.~G.}, {\sc and}
  {\sc Hermenegildo, M.~V.} 1999.
\newblock Automatic compile-time parallelization of logic programs for
  restricted, goal level, independent {AND}-parallelism.
\newblock {\em Journal of Logic Programming\/}~{\em 38,\/}~2, 165--218.

\bibitem[\protect\citeauthoryear{Pontelli, Gupta, Pulvirenti, and
  Ferro}{Pontelli et~al\mbox{.}}{1997}]{Pontelli97automaticcompile-time}
{\sc Pontelli, E.}, {\sc Gupta, G.}, {\sc Pulvirenti, F.}, {\sc and} {\sc
  Ferro, A.} 1997.
\newblock Automatic compile-time parallelization of prolog programs for
  dependent and-parallelism.
\newblock In {\em Proceedings of the 14th International Conference on Logic
  Programming}. Leuven, Belgium, 108--122.

\bibitem[\protect\citeauthoryear{Shen, Costa, and King}{Shen
  et~al\mbox{.}}{1998}]{shen_98_granularity-control}
{\sc Shen, K.}, {\sc Costa, V.~S.}, {\sc and} {\sc King, A.} 1998.
\newblock Distance: a new metric for controlling granularity for parallel
  execution.
\newblock In {\em Proceedings of the 1998 joint international conference and
  symposium on Logic programming}. MIT Press, Cambridge, MA, USA, 85--99.

\bibitem[\protect\citeauthoryear{Somogyi, Henderson, and Conway}{Somogyi
  et~al\mbox{.}}{1996}]{jlp}
{\sc Somogyi, Z.}, {\sc Henderson, F.}, {\sc and} {\sc Conway, T.} 1996.
\newblock The execution algorithm of {Mercury}, an efficient purely declarative
  logic programming language.
\newblock {\em Journal of Logic Programming\/}~{\em 26,\/}~1-3
  (October-December), 17--64.

\bibitem[\protect\citeauthoryear{Tannier}{Tannier}{2007}]{tannier:2007:paralle%
l-mercury}
{\sc Tannier, J.} 2007.
\newblock Parallel {Mercury}.
\newblock M.S.\ thesis, Institut d'informatique, Facult\'es Universitaires
  Notre-Dame de la Paix, 21, rue Grandgagnage, B-5000 Namur, Belgium.

\bibitem[\protect\citeauthoryear{von Praun, Ceze, and Ca\c{s}caval}{von Praun
  et~al\mbox{.}}{2007}]{von_Praun:2007:implicit_parallelism_with_ordered_trans%
actions}
{\sc von Praun, C.}, {\sc Ceze, L.}, {\sc and} {\sc Ca\c{s}caval, C.} 2007.
\newblock Implicit parallelism with ordered transactions.
\newblock In {\em Proceedings of the 12th Symposium on Principles and Practice
  of Parallel Programming}. San Jose, California, 79--89.

\bibitem[\protect\citeauthoryear{Wang and Somogyi}{Wang and
  Somogyi}{2011}]{wang_dep_par_conj}
{\sc Wang, P.} {\sc and} {\sc Somogyi, Z.} 2011.
\newblock Minimizing the overheads of dependent {AND}-parallelism.
\newblock In {\em Proceedings of the 27th International Conference on Logic
  Programming}. Lexington, Kentucky.

\end{thebibliography}

\tr{
\newpage

\begin{table*}
\begin{tabular}{lll}
Item & Owner & Status \\
\hline

\todoitem{
when calculating the average time of consumption of a variable
in a branched structure,
try giving ZERO weight to arms that themselves are trivial in cost.
example: by\_phases in the front end.
}{n/a}{later} \\

\todoitem{
fix the report generated by mdprof\_feedback:
it says it reports the number of parallelized conjunctions,
when what it reports is the number of procedures
that CONTAIN parallelized conjunctions.
If some procedures contain more than once,
the report is misleading.}{pbone}{done} \\

\todoitem{
implement option (default on) to respect boundaries
when calculating variable signal and wait times:}{pbone}{DONE} \\

\todoitem{
detecting parallelism opportunities involving nonatomic
conjuncts}{pbone}{DONE} \\

\todoitem{
implementation of throttling by default}{zs and pbone}{DONE} \\

\todoitem{
branch-and-bound on all conjuncts,
then peel off conjuncts at start and end}{zs and pbone}{DONE} \\

\todoitem{post message about Boehm parameters}{pbone}{DONE} \\

\todoitem{
consider keeping a non-first parallel conjunct on the original
engine}{all}{Different paper} \\

\todoitem{specialization: parallel for some ancestors, not others}{zs, with
feedback work by pbone}{Different paper} \\

\todoitem{
grouping conjuncts that together do one thing, reduces $N$ for
B'n'B}{pbone}{low priority} \\

\todoitem{best-first, not depth-first search}{all}{low priority} \\

\todoitem{gzip Deep.data}{zs}{low priority} \\

\todoitem{include Deep.procrep in Deep.data}{zs}{low priority} \\

\todoitem{
Push costly calls into branching code, analysis part}{pbone}{DONE} \\

\todoitem{
Push costly calls into branching code, compiler part}{zs}{DONE} \\

\todoitem{
  Investigate Haskell's nofib benchmark for useful
  programs.}{peters}{DONE} \\

\todoitem{
  Investigate the Mars compiler as a benchmark.}{peters}{} \\

\todoitem{
  Introduce goal IDs to replace goal paths in many places}{zs}{DONE} \\

\todoitem{
  Fix --max-contexts-per-thread}{pbone}{} \\

\todoitem{
  Put workstealing deques in engines, not contexts}{pbone}{DONE} \\

\todoitem{
enhance work stealing and stack segment caching algorithms}{zs
  and pbone and maybe wangp}{} \\

\todoitem{
finish deep profiler paper}{zs}{} \\

\todoitem{
finish dep and parallel paper}{zs}{WIP} \\

\todoitem{
Mission Critical benchmark}{pbone and probably zs}{} \\

\todoitem{compiler as benchmark}{all}{} \\

\todoitem{description of the cost of recursive calls algorithm}{all}{}
\\

\todoitem{
divide and conquer limit:
same seq vs parallel for double recursive procs}{all}{} \\

\todoitem{consider how recursive calls in parallelized conjunctions
  affect the speedup}{pbone with help}{} \\

\todoitem{throttle only looping/D'n'C code}{pbone}{} \\

\todoitem{Commit various things, benchmarks, stats program to some
  repository}{pbone}{DONE} \\

\todoitem{promise that you don't care what random number sequence you get}{}{}

\end{tabular}
\end{table*}
}

\reviewcomments{
\newpage
\appendix
\section*{Changes since original submission}

Here we list all the reviewer's comments,
and our responses to those comments.

Reviewer 1:
\emph{It is insufficient
to consider each conjunction on its own without consider any further
parallelism within the execution of the goals in the conjunction.}

\textbf{We now address this point towards the end of section 3.}

Reviewer 1 and 2:
\emph{More results, and more comparisons.}

\textbf{We now report some more benchmark results.
We have many more results, but no room to discuss them.}

Reviewer 1:
\emph{The choice of automatic parallelisation \ldots the programmer
might know where parallelism is most profitable.}

\textbf{The Mercury system allows the programmer to ask for a conjunction
to be executed in parallel simply by using ampersand instead of comma
as the conjunction operator.
The recommendations of our feedback tool do not override
the programmer's explicit choice; we now say so explicitly.}

\textbf{
However, in our experience, many of the places where we \emph{thought}
we could parallelize our own programs
turned out not to be \emph{profitably} parallelizable,
due to dependencies that we had forgotten about.
}

Reviewer 1:
\emph{Consider the case \texttt{P :- Q \& P}. then P is very likely
to do much more computation than Q, so the overlap (ignoring any
dependencies) of Q will only be a small fraction of P, but because of the
recursive calls to the parallel conjunction, this is a good candidate for
parallelisation.}

\textbf{We make this point
in the third last paragraph of section 4.} 

Reviewer 1:
\emph{Show that the automatic parallelisation does better than
running all 'reasonable' goals in parallel.  Show that considering the
timing of producer/consumer is effective in producing better
speedups.}

\textbf{So far, we have only suggestive
(as opposed to convincing) data for this.
We have added discussion about this to the performance section.} 

Reviewer 1:
\emph{Some information/results on how the
predicted speedups compare to the real speedups you get, and
also on the cost of doing the automatic parallelisation.}

\textbf{
While our feedback tool predicts the local speedup from parallelizing a single conjunction,
it does not (yet) have code to predict an overall speedup for the whole program.
We intend to add such code in the near future.
The problem is a bit more complex than it appears:
the overall speedup is not just the sum of the effects of the local speedups,
because the conjunctions that we want parallelize in one context
we may not want to parallelize in other contexts.
This is not an issue in our small benchmarks,
but it is an issue in real programs.
We propose a solution as future work at the end of the conclusion.
}

\textbf{
The cost of doing the automatic parallelisation is negligible,
and we now report that fact.}

Reviewer 1:
\emph{``goals expensive enough to be worth executing in parallel are
rarely independent'' is a very strong statement, and you give no support
for it.}

\textbf{We have now added an example,
which we hope the reviewer accepts as sufficient:
in the Mercury compiler itself,
there are 53 conjunctions containing two or more expensive goals,
but in only one of those conjunctions are the expensive goals independent.
The numbers for other Mercury programs may be different,
but in our experience, the same conclusion holds for them too.}

Reviewer 1:
\emph{Is the parallelism described in Section~\ref{sec:backpar} new?
It is not very clear from the material presented here,
and no references are given.}

\textbf{That work took place before this one,
but it is described for the first time in another paper at ICLP 2011;
we have added a reference to that paper.}

Reviewer 1:
\emph{What happens if the producer produces a value, but
which is then backtracked over? Restricting a goal to have only one
solution does not prevent localized backtracking.}

\textbf{It does not, but in Mercury,
code inside a det computation that can be backtracked over
(such as the condition of an if-then-else)
can never make bindings that are visible from outside that det computation.
The situation that question asks about can never arise,
and now we make a parenthetical comment to this effect.
This is by design; we have known that this is an issue since the 1980s.
}

Reviewer 1:
\emph{It is not clear that the deterministic conjunct restriction is
``not significant'' as you state, simply because most procedure are
deterministic. The same can be said of Prolog predicates -- that
most of them are deterministic, and I think the work in \&-Prolog
(independent and-parallelism) and DDAS (dependent and-parallelism)
suggest that this deterministic restriction does reduce
parallelism.}

\textbf{It does reduce the number of clauses whose parallelization is possible,
but systems like DDAS can handle parallel execution of nondeterministic code
only at the cost of incurring huge overheads,
most of which are paid even when the conjuncts are deterministic.
We would prefer to parallelize 80\% of a program with 5\% or 10\% overhead,
instead of parallelizing 100\% of the program with 200\% or 400\% overhead,
and now we say so.
(We don't think those high overhead figures ever appeared in published papers
for systems like DDAS, but they are derivable from timings published for them
for standard benchmarks, and timings published for fast sequential systems like
SICStus on the same hardware for the same benchmarks. They are also rather
obvious if one compares the complexity of the basic operations in systems like
DDAS with the complexity of the corresponding operations in the WAM.
And the WAM operations themselves are much more complex
than the corresponding operations in the Mercury implementation.}

\textbf{We have a script that prints statistics about Mercury programs,
including what fraction of the predicates have each determinism.
In most Mercury programs, at least 75\% of predicates are det,
most of the rest are semidet, with only 1 to 5\% being nondet or multi.
Even then, the nondet and multi predicates tend to have small runtimes.
The reason for this is that the standard SLD search algorithm
used by both Prolog and Mercury is very naive,
and thus it is not efficient at searching large spaces.
If the search space is large, programmers tend to prefer 
to program the search themselves,
and the code they write for doing this
is usually mostly det code, with some semidet tests.
The G12 project in our department
uses Mercury to program constraint solvers which search enormous spaces,
and those solvers are certainly written this way.}

Reviewer 1:
\emph{The restriction that a shared variable must be instantiated
to a ground value seems to further restrict parallelism. In exploiting
dependent and-parallelism in Prolog, significant parallelism comes from
the producer incrementally producing the elements of a shared variable
(e.g. each element of a list).}

\textbf{That is correct. However, it is a limitation
of the current mechanisms for implementing dependent parallelism in Mercury.
The system that this paper describes
only generates recommendations for where to apply those mechanisms,
and thus it cannot transcend the limitations of those mechanisms.
When the implementation of dependent parallelism in Mercury
is improved to allow such cooperation between producers and consumers,
we intend to modify the feedback tool to take advantage of that capability.
The task of modifying the underlying system is already on our todo list.
It is relatively easy to do if one is willing to accept high overheads;
doing it with low overheads, which is what we want to do,
is quite challenging.}

Reviewer 1:
\emph{4 is a relatively low number of CPUs, so it should still be
possible to get closer to a speedup of 4.}

\textbf{
Many parallel Prolog systems have primitive operations
that do quite significant amounts of work (dozens of instructions)
for simple tasks like binding a variable.
Many of those instructions access only virtual machine data,
which is very likely to be hot data
and thus very unlikely to cause cache misses.
Parallel Mercury programs have MUCH lower overheads;
most bindings bind variables that are not shared between parallel conjuncts,
and those bindings need only a single instruction.
This instruction will access memory
that is not part of the implementation of the underlying virtual machine,
and is therefore significantly less likely to be in the cache.
Parallel Mercury programs therefore put
significantly more pressure on the shared bus to memory
than parallel Prolog programs do.
Given these facts, we think a speedup of 3.87 (was 3.76) on 4 CPUs
is pretty close to what one can realistically hope for.
For the other benchmarks, the speedup we can hope for is even lower.
We have added to the performance section
as thorough a discussion of this issue as we have room for.
}

Reviewer 1:
\emph{``continue execution after a parallel conjunction on a
different CPU from the one before the conjunction.''
Will the continuation be done by the CPU that finished the last goal in
the conjunction? If so, I am not sure why this should be significantly
worse that the one before the conjunction, because executing the
goal in the conjunction should "warm" the cache.}

\textbf{The problem was that often, the "if so" was not so:
the CPU that executed the code after the conjunction
switched to that task from executing code from another conjunction.
We have now updated our implementation to address the problem.
Now, the engine that last finishes a conjunct of a parallel conjunction
(whichever conjunct that happens to be)
will switch to executing the code after the conjunction
to take advantage of its warm cache.
As we mentioned, it was always our intention to do this;
we were prevented only by some technical problems
that we did not have time to fix before our initial submission.
}

Reviewer 1:
\emph{Section 7 "Conclusion", first paragraph: ``Most previous
work...focused on granularity control....Our feedback tool  show that
this is far from enough.''  Most of the Prolog related granularity
control work was designed for independent
and-parallelism, so the issue of producer-consumer dependencies is not an
issue.}

\textbf{We have reworded that paragraph to avoid that implication.}

Reviewer 1:
\emph{Further, (Shen et al 1999) does not do granularity control as you
described. It actually was suggesting that a different metric than the
size of the goal should be used to improve the efficiency of independent
and-parallel execution, because of recursive spawning overheads.}

\textbf{We have reworded the description of the cited paper.}

Reviewer 1:
\emph{Your point that granularity control is not enough for your
parallel model will be stronger if you can show results that demonstrate
this, i.e. the speedup being much lower if you parallelize goal with only
granularity control.}

\textbf{This is a repeat of a point above, and it is addressed above.}

Reviewer 2: \emph{
It is also not clear whether the producer-consumer relation is just an
artifice: if the consumer does most of the execution
without waiting for the producer, then maybe it does not care so much.
In this case, we would be closer to NSIAP, where you use the futures
as a kind of replacement for the ground term.}

\textbf{We do not understand what the reviewer means by ``artifice''.
Regardless of how much work the consumer can do
before it needs the value of the shared variable,
the presence of a shared variable makes the two conjuncts dependent.
If the consumer can do lots of work before it needs the value of the
shared variable, that is nice; that is exactly the situation we look for,
since it allows large overlap and hence a large speedup. However,
the underlying implementation must still handle the dependency correctly,
and our system must still try to estimate the overlap as accurately as it can.
}

\textbf{
NSIAP, Non-Strict Independent And-Parallelism, is a self-contradictory term,
since the ``non-strict'' part means that
the conjuncts may actually be dependent,
as long as they are dependent in a particular way.
The restrictions that NSIAP imposes on dependencies
are guaranteed to be met by the Mercury mode system.
We hope this clarifies the relationship of our work with NSIAP,
because we do not know what else the comment is asking for.
}

\textbf{
A future is not a \emph{replacement} for a ground term;
it is a \emph{placeholder} for a ground term,
from which the ground term can be extracted when it becomes available.
}

\emph{You really must discuss related work
in order to understand these issues. }

\textbf{This paper is about a system
for generating recommendations about what to parallelize,
not about the underlying system that implements those recommendations.
(The paper by Wang and Somogyi is about that underlying system.)
Given that we have limited room for discussing related work,
we have chosen to devote the bulk of it
to work that is also about choosing what to parallelize.
We do still talk about other work related to the underlying system,
even though most of it is not relevant in our context;
as we note in the paper,
most of the work on the implementation of AND-parallelism for Prolog
is not easily compared with AND-parallelism in Mercury,
because most of the key problems for Prolog do not arise for Mercury.
We do not have room for a detailed explanation
of \emph{why} these problems do not arise for Mercury.}

Reviewer 2: \emph{
Page 1: the best parallelization opportunities appear: why?}

\textbf{The quoted sentence says
``the best parallelization opportunities occur
where two goals take a significant and roughly similar time to execute''.
The ``significant'' part is explained in the next sentence.
The ``roughly similar'' part is a simple application of Amdahl's law:
if the two conjuncts take time t and 100t respectively,
then the best possible improvement is only from 101t to 100t.}

Reviewer 2: \emph{
Page 2 has an interesting discussion
about the sources of parallelism in Mercury.
The problem is that the discussion is not supported by much evidence.}

\textbf{It is something one picks up
if one reads (or writes) enough Mercury code.
We now give some evidence from such a program in the conclusion.}

Reviewer 2: \emph{
By Indep And-parallelism do you also refer to NSIAP?
For example \texttt{map\_fold} looks like a typical example of NSIAP.}

\textbf{No. As we mention above, despite the origin of the acronym,
NSIAP is actually a form of dependent AND-parallelism.
As we discuss in the text,
the relevant conjunction in \texttt{map\_fold} is dependent, not independent.
NSIAP can handle it only because it can handle some forms of dependencies.}

Reviewer 2: \emph{
What is the connection between futures and suspension records
used in the CCLs?}

\textbf{That depends on what kinds of suspension records you are talking about.
In the earliest committed choice languages,
it was possible for a goal to suspend
until any one of several variables became bound.
The suspension records used in such systems were significantly more complex,
and operations on them were significantly more time consuming,
than operations on futures.
Later, more restrictions were imposed on the source languages
so that implementation could use lower-overhead forms of synchronization.
However, none were faster than futures,
and (due to the absence of mode information,
even in Parlog, which lost the mode system it started out with),
they had to be used for FAR more variable accesses,
and thus incurred far higher overheads.}

\textbf{
We believe futures are as simple and as quick a mechanism
for handover from producer to consumer as one can imagine.
We don't believe there is any way
to build anything faster in software using current hardware.
In the 1980s, Iannucci proposed I-structures,
a paper design for the implementation of futures in hardware.
ICOT in Japan may have actually built hardware based on such designs,
but if so, they sank without a trace.
}

Reviewer 2: \emph{
Page 3: You really should refer to Reform Prolog in page 3.
Reform took these ideas much further.}

\textbf{Yes, it did. We have added a reference.}

Reviewer 2: \emph{
Page 5: there is little discussion on item 2:
is your analysis valid if one changes the scale/size of the problem?}

\textbf{We have added some discussion of this point.}

Reviewer 2: \emph{
page 7: I didn't follow the problem with the high-level call}

\textbf{If the question is about higher order calls:
the Mercury compiler does not currently track
the set of predicates that a higher order variable can refer to.
Even if it did, the callee could be
a predicate defined in another module of the program,
in which case \emph{this} compiler invocation would not have access to it.}

Reviewer 2: \emph{
page 9 \texttt{Var\_ij}:  why the underscore?}

\textbf{Because without it, readers would find it harder to know
where the array name ends and the index values begin,
and because the algorithms look ugly when using the math font.}

Reviewer 2: \emph{
Do you have significant variance in the results?
Did you call the garbage collector?}

\textbf{Yes, there was variance in the results, and we now say so explicitly.
We don't report details of this variance,
since it would complicate and widen the table,
and it is not usually included with benchmarks.
Yes, mandelbrot and raytracer both cause several garbage collections, 
and we now mention that fact.
Matrixmult does not invoke gc;
it allocates memory only for the two input matrices
and the output matrix, all in the form of lists of lists.}

\textbf{We intend to publish on the Mercury web page, next to this paper,
a package that anyone can download for reproducing our benchmark results.}

Reviewer 2: \emph{
Page 14: actually, linear speedups have been reported for LP
(yapor/thor paper) and are often found in the literature.
Cache effects may also result in super linear effects,
if there is little communication. }

\textbf{Yes, superlinear speedups are possible
both because the parallel version of a program
necessarily executes operations in a different order
than the sequential version,
and because it has access to the caches of several CPUs, not just one.
However, there are still many bottlenecks in real CPUs;
the connection to memory and stop-the-world garbage collection
(which we mention in the paper) are just two.
Then there are the costs of parallelism, which we list in the paper.
Thus we believe it is unrealistic to expect linear speedups
while maintaining low overheads.}

Reviewer 3: \emph{
1) It does not connect with the large body of research in parallel
execution in Logic Programming --- and note I do \emph{not} mean Concurrent
Logic Programming.  In fact, some of the usual general assumptions in
parallel LP are not valid (e.g., the type and mode system make up for
a large part of the work in type / mode inference), and the
restrictions on the terms and instantiation states change somewhat the
landscape.  These should be highlighted.}

\textbf{It changes the landscape \emph{entirely}.
Many problems in the parallel execution of Prolog and Prolog-like languages,
like testing the independence of goals
in systems that support only independent AND-parallelism,
discovering producer-consumer relationships at runtime
in systems that also support dependent AND-parallelism,
and having to handle nondeterministic conjuncts,
disappear completely,
with the answers to the problem being presented on a silver platter
by the semantic analysis passes of the Mercury compiler.
Our group designed Mercury specifically to ensure this.
We now say mention this in the conclusion.
However, this is a difference in the underlying system,
not in the auto-parallelization system,
so it is not the main topic of the paper,
which is why the mention does not include extensive discussion.
}

Reviewer 3: \emph{
2) Related to the previous point, it needs to refer to and compare
with previous work on parallelization, since, besides using profiling
to decide and drive the annotation algorithm, a lot of the concerns
are shared.  A classical references is
\cite{DBLP:journals/jlp/MuthukumarBBH99}
and two more recent papers are
\cite{DBLP:journals/tcs/GrasH09,DBLP:conf/lopstr/CasasCH07}.}

\textbf{We now have more discussion of related work,
including those three papers.
Unfortunately, we don't have room to have as extensive a discussion
as we and the reviewers would like.}

Reviewer 3: \emph{
3) The experimental performance assessment needs improvement.  On one
hand one has to check any parallel execution framework with the
simple, well-known examples where the expected behavior is known:
matrix multiplication, Fibonacci, etc. where parallelism is
independent and speedups are expected to be linear --- in order to
know what is the behavior of the system in optimal conditions.  And,
in any case, details on the benchmarks are necessary.  For example, in
the case of the raytracer I would have expected an almost linear
speedup, since one can assign a screen pixel (or set thereof) to an
independent computation.  Is it done like that, or is there something
which disallows this simple assignment?  Small speedups do not tell
us whether there is a problem with the parallelization, the parallel
engine, or if there is simply no more parallelism available in the
program.}

\textbf{We have added results for matrix multiplication;
we don't have room to report on more benchmarks.
We now report that the raytracer's bottleneck is the garbage collector,
which we don't have any effective control over.
(We use the Boehm-Demers-Weiser collector.
While it works very well, its internals are very complex
and almost completely undocumented,
so modifying it is extremely hard.)}

Reviewer 3: \emph{
What you do is binding-level parallelism as opposed to goal-level
parallelism which has being mainstream (but not exclusive) in
Prolog-like languages.  You may want to point this out in page 2,
para. 2.}

\textbf{Since ``binding-level parallelism'' is not (yet) standard terminology,
we do not think that would be very helpful.}

Reviewer 3: \emph{
How do you determine the ``representative'' input you initially feed the
program with?  Some programs are extremely sensitive to the input
(e.g., you mention QuickSort, which is one of them, but dismiss them
it because you claim it is a very unlikely pattern).  }

\textbf{It is up to the user of our system to select representative input,
and if the profiling input is not representative,
then users cannot expect
the results of any profile-directed program transformations to be optimal.
This is par for the course for any profiling-based program transformation,
and we now point it out explicitly.}

\textbf{We do not mention anything about the sensitivity of quicksort
in this paper.}

Reviewer 3: \emph{
Related to the previous point, you also mention that your system can
predict speedups for virtually all candidates for parallelization.  I
am afraid I do not see any proof or support in the paper.}

\textbf{We make that mention in the context of computing speedups
depending on the recursion structure of the procedure involved.}

\textbf{
Our feedback tool processes
the representations of the programs it is working on.
We added some code to it to classify
all directly or mutually recursive predicates into three categories:
single recursion (like map), double recursion (like quicksort),
and irregular recursion (everything else).
In every program we have run the tool on,
the number of predicates in the first two classes vastly outnumbered
the predicates in the third class.
For small programs, the third class was usually empty.
We now mention in the paper
that our assertion is based on hard data.}

Reviewer 3: \emph{
Moreover,
you are implicitly giving up on irregular computations whose shape are
highly dependent on the input data (e.g., protein matching /
folding?).  What do you if you are in the case where you cannot
predict speedup?  Which parallelization do you choose?}

\textbf{As mentioned for a previous comment by the same reviewer:
it is up to the user of our system to select representative input,
and if the profiling input is not representative,
our results cannot be expected to be optimal.
However, a single parallelization can be a good idea
for many different shapes of computation,
while being a bad idea for other shapes.
The important thing is not whether
the shape of the computation during the profiling run is the same as
the shape of the computation during the actual run,
but whether the parallelization recommendation generated by the profiling tool
generates useful speedups or not.
The parallelization can be right even if the shape is not the same.
We now mention this in the body of the paper.}

\textbf{In cases where we cannot predict speedup because
the recursion structure of the recursive predicate(s) involved is irregular,
we do the conservative thing and choose not to parallelize.
Changing this is future work.}

Reviewer 3: \emph{
When you write ``The condition of this if-then-else will test at
runtime whether spawning off a new job is a good idea or not.'', how do
you decide whether it is or not a good idea?}

\textbf{This question is answered in the body of the paper
a few lines after the quoted sentence.}

Reviewer 3: \emph{
It would be interesting to know which are the speedups predicted with
the testing / profiling input date \emph{and} which are the speedups
attained using precisely that data (besides speedups obtained with
other data).}

\textbf{We have run the benchmarks with the training data as well as different data.
In the paper, we report results for the usual case of
the test data being different from the training data.
However, we get similar speedups for the training data itself.
we plan to include that in a future longer version of the paper.}

Reviewer 3: \emph{
``Currently, the Mercury runtime system often continues execution, on
completion of a parallel conjunction, on a CPU different from the one
being used before that parallel conjunction''.  You should change that
as it can be a source of inefficiency, especially in the case of
recursive predicates.}

\textbf{We have changed that.
See our response to Reviewer \#1, who made the same point.}
}

\end{document}